\documentclass[sigconf,authorversion]{acmart}


\usepackage{ifthen}
\newboolean{draftversion}

\setboolean{draftversion}{false} 

\ifthenelse{\boolean{draftversion}}{\setlength{\paperwidth}{10.5in}\setlength{\marginparwidth}{1in}\setlength{\hoffset}{1in}}{}
\newcommand{\jiska}[1]{\ifthenelse{\boolean{draftversion}}{{\color{blue}{\textbar}\hspace{-.4em}\marginpar{\color{blue}{\emph{JC:}\\#1}}}}{}}
\newcommand{\ms}[1]{\ifthenelse{\boolean{draftversion}}{{\color{red}{\textbar}\hspace{-.4em}\marginpar{\color{red}{\emph{MS:}\\#1}}}}{}}
\newcommand{\mh}[1]{\ifthenelse{\boolean{draftversion}}{{\color{OliveGreen}{\textbar}\hspace{-.4em}\marginpar{\color{OliveGreen}{\emph{MH:}\\#1}}}}{}}

\newcommand*\phantomas[3][c]{%
\ifmmode 
\makebox[\widthof{$#2$}][#1]{$#3$}%
\else 
\makebox[\widthof{#2}][#1]{#3}%
\fi 
}

\newif\ifblinded
\blindedfalse

\blindedfalse

\usepackage{placeins} 

\usepackage[colorinlistoftodos,prependcaption,disable]{todonotes} 
\presetkeys%
    {todonotes}%
    {inline,backgroundcolor=black!20}{}

\usepackage{amsmath}

\usepackage{xspace}

\usepackage{flushend}

\usepackage{tikz}
\usetikzlibrary{chains}
\usetikzlibrary{positioning}
\usetikzlibrary{shapes.geometric}
\usetikzlibrary{decorations.pathreplacing}
\usepackage{pgfplots}
\usepackage{capt-of}
\usepackage{graphicx}
\usepackage{caption}
\usepackage{booktabs}
\usepackage{subcaption}
\newlength\figureheight
\newlength\figurewidth

\usepackage{tabularx}

\newcolumntype{b}{X}
\newcolumntype{s}{>{\hsize=.5\hsize}X}

\usepackage[binary-units=true]{siunitx}

\usepackage{listings} 
\usepackage[nolist]{acronym}
\begin{acronym}
\acro{AOSP}{Android Open Source Project}
\acro{AWS}{Amazon Web Services}
\acro{JVM}{Java Virtual Machine}
\acro{ADB}{Android Debug Bridge}
\acro{HCI}{Host Controller Interface}
\acro{RAM}{Random Access Memory}
\acro{ROM}{Read Only Memory}
\acro{IoT}{Internet of Things}
\acro{LNA}{Low Noise Amplifier}
\acro{AGC}{Automatic Gain Control}
\acro{UART}{Universal Asynchronous Receiver-Transmitter}
\acro{TLV}{Type-Length-Value}
\acro{EEPROM}{Electrically Erasable Programmable Read-Only Memory}
\acro{UI}{User Interface}
\acro{RNG}{Random Number Generator}
\acro{NIST}{National Institute for Standards and Technology}
\acro{IO}{Input and Output}
\acro{LM}{Link Manager}
\acro{LMP}{Link Manager Protocol}
\acro{WEP}{Wired Equivalent Privacy}
\acro{SIG}{Special Interest Group}
\acro{BR}{Basic Rate}
\acro{EDR}{Enhanced Data Rate}
\acro{LE}{Low Energy}
\acro{BLE}{Bluetooth Low Energy}
\acro{TDD}{Time-Division Duplex}
\acro{AFH}{Adaptive Frequency Hopping}
\acro{L2CAP}{Logical Link Control and Adaption Protocol}
\acro{PDU}{Program Data Unit}
\acro{USB}{Universal Serial Bus}
\acro{RFCOMM}{Radio Frequency Communications}
\acro{SDP}{Service Discovery Protocol}
\acro{TCP}{Transmission Control Protocol}
\acro{QoS}{Quality of Service}
\acro{HEC}{Header-Error-Check}
\acro{LLID}{Logical Link ID}
\acro{CRC}{Cyclic Redundancy Check}
\acro{MIC}{Message Integrity Check}
\acro{PCM}{Pulse-code Modulation}
\acro{AHB}{Advanced High-performance Bus}
\acro{LSB}{Least Significant Bit}
\acro{TRM}{Technical Reference Manual}
\acro{IVT}{Interrupt Vector Table}
\acro{NVIC}{Nested Vectored Interrupt Controller}
\acro{NMI}{Non-maskable Interrupt}
\acro{FPB}{Flash Patch and Breakpoint}
\acro{SDIO}{Secure Digital Input Output}
\acro{ACL}{Asynchronous Connection-Less}
\acro{SCO}{Synchronous Connection-Orientated}
\acro{SP}{Stack Pointer}
\acro{FIFO}{First-In-First-Out}
\acro{CLI}{Command Line Interface}
\acro{API}{Application Programming Interface}
\acro{UAP}{Upper Address Part}
\acro{LAP}{Lower Address Part}
\acro{SDR}{Software-Defined Radio}
\acro{RF}{Radio Frequency}
\acro{DMA}{Direct Memory Access}
\acro{TID}{Transaction ID}
\acro{SSP}{Secure Simple Pairing}
\acro{TLS}{Transport Layer Security}
\acro{JW}{Just-Works}
\acro{NC}{Numeric-Comparison}
\acro{OOB}{Out-Of-Band}
\acro{MITM}{Man-In-The-Middle}
\acro{IP}{Internet Protocol}
\acro{PAN}{Personal Area Network}
\acro{ECDH}{Elliptic Curve Diffie-Hellman}
\acro{AES}{Advanced Encryption Standard}
\acro{CCM}{Counter mode with CBC-MAC}
\acro{ISM}{Industrial, Scientific and Medical}
\acro{LE SC}{LE Secure Connections}
\acro{PHY}{Physical Layer}
\acro{GATT}{Generic Attribute Profile}
\acro{MAC}{Media Access Control}
\acro{AFH}{Adaptive Frequency Hopping}
\acro{AoA}{Angle of Arrival}
\acro{AoD}{Angle of Departure}
\acro{AWDL}{Apple Wireless Direct Link}
\acro{RSSI}{Received Signal Strength Indicator}
\acro{BPCS}{Broadcom Proprietary Control Signaling}
\acrodefplural{SDR}[SDRs]{Software-defined Radios}
\acrodefplural{LM}[LMs]{Link Managers}
\acrodefplural{CLI}[CLIs]{Command Line Interfaces}

\end{acronym}

\copyrightyear{2019}
\acmYear{2019}
\setcopyright{acmlicensed}
\acmConference[MobiSys '19]{The 17th Annual International Conference on Mobile Systems, Applications, and Services}{June 17--21, 2019}{Seoul, Republic of Korea}
\acmBooktitle{The 17th Annual International Conference on Mobile Systems, Applications, and Services (MobiSys '19), June 17--21, 2019, Seoul, Republic of Korea}
\acmPrice{15.00}
\acmDOI{10.1145/3307334.3326089}
\acmISBN{978-1-4503-6661-8/19/06}

\begin{document}

\title[InternalBlue]{InternalBlue -- Bluetooth Binary Patching and \\Experimentation Framework}

\newcommand{\mytool}{\emph{InternalBlue}\xspace}
\renewcommand{\hex}[1]{{\texttt{0x#1}}}

\ifblinded
\numberofauthors{1}
\author{
Authors blinded for review.
}

\else

\author{Dennis Mantz }
\email{dmantz@seemoo.de}
\affiliation{%
  \institution{TU Darmstadt, Secure Mobile Networking Lab}
 \city{Darmstadt}
 \country{Germany}
}
\author{Jiska Classen}
\email{jclassen@seemoo.de}
\affiliation{%
  \institution{TU Darmstadt, Secure Mobile Networking Lab}
 \city{Darmstadt}
 \country{Germany}
}
\author{Matthias Schulz}
\email{mschulz@seemoo.de}
\affiliation{%
  \institution{TU Darmstadt, Secure Mobile Networking Lab}
 \city{Darmstadt}
 \country{Germany}
}
\author{Matthias Hollick}
\email{mhollick@seemoo.de}
\affiliation{%
  \institution{TU Darmstadt, Secure Mobile Networking Lab}
 \city{Darmstadt}
 \country{Germany}
}

\fi

\begin{abstract}
Bluetooth is one of the most established technologies for short range digital
wireless data transmission. With the advent of wearables and the
\ac{IoT}, Bluetooth has again gained importance, which makes security research and protocol optimizations imperative.
Surprisingly, there is a lack of openly available tools and experimental
platforms to scrutinize Bluetooth. In particular, system aspects and close to hardware protocol layers are mostly uncovered.

We reverse engineer multiple \emph{Broadcom} Bluetooth chipsets that are widespread in off-the-shelf devices. 
Thus, we offer deep insights into the internal architecture of a popular commercial family of Bluetooth controllers used in smartphones, wearables, and IoT platforms. 
Reverse engineered functions can then be altered with our \mytool \emph{Python} framework---outperforming evaluation kits, which are limited to documented and vendor-defined functions.
The modified Bluetooth stack remains fully functional
and high-performance. Hence, it provides a portable low-cost research platform.

\mytool is a versatile framework and we demonstrate its abilities by implementing tests and demos for known Bluetooth vulnerabilities. Moreover, we discover a novel critical security issue affecting a large selection of \emph{Broadcom} chipsets that allows executing code within the attacked Bluetooth firmware. We further show how to use our framework to fix bugs in chipsets out of vendor support and how to add new security features to Bluetooth firmware.


\vspace*{5em} 

\end{abstract}

\maketitle

\todo{maybe add in a positive way what our abilities are (extensible, go beyond vendor defined ...)}






%



\section{Introduction}

\todo[color=green!20]{\emph{InternalBlue has the potential to spark greater engagement with Bluetooth.}

\emph{this paper does add to our understanding of the Bluetooth ecosystem.}

Adapt some of this positive feedback into the motivation.
}

Bluetooth, \emph{the} standard for wireless short range communication, has been
around for almost 25 years since Ericsson developed it in 1994. In the
early days it was mainly applied to wireless headphones, hands-free speakerphones,
and replacement of infrared data links between devices~\cite{bluetooth-history}. Today, Bluetooth
experiences a comeback with the use of wearables and the
\ac{IoT}, often using \ac{BLE} introduced in version 4.0. Its latest specifications, Bluetooth 5.0 and 5.1, provide new
interesting features such as mesh networking and localization \cite{2017:SIG, 2019:SIG}, and indicate that
Bluetooth will play an important role in the future of wireless communication.



Bluetooth security and performance have only been studied selectively, which is a stark contrast to the extensive analysis of the Wi-Fi standard over the same period of time. This can be partially attributed to the availability of powerful, open-source tools
which allow easy experiments on raw Wi-Fi frames with low-cost, off-the-shelf
hardware. When the first patches for Wi-Fi drivers enabled the  so-called
\emph{monitor mode} and frame injection capabilities,
researchers soon implemented practical attacks on low-level parts of the Wi-Fi stack and the currently deprecated \ac{WEP} standard~\cite{2001:fms, 2006:bittau} in short order.
Only recently~\cite{2017:googleprojectzero, 2017:artenstein, quarkslab}, the firmware
running on Wi-Fi cards has been shown to contain severe over-the-air vulnerabilities. 
Publicly available tools allow to alter off-the-shelf \emph{Broadcom} Wi-Fi cards~\cite{nexmon:project}, along with easy to understand open source Wi-Fi SDR implementations~\cite{warp:project, gr-ieee802-11:project}.

\emph{Blueborne}
is a state-of-the-art collection of weaknesses uncovered in most of the major Bluetooth stacks, and has raised awareness of different issues concerning Bluetooth security. 
However, \emph{Blueborne} targets host-side Bluetooth drivers in
opposition to the lower layers of the protocol, below the \ac{HCI},  which are handled in firmware
and are still difficult to audit.
A recent Bluetooth attack concerns \ac{ECDH} key exchange used during device pairing, where the attacker replaces public key coordinates during transmission, which were not checked in most implementations~\cite{2018:biham}.

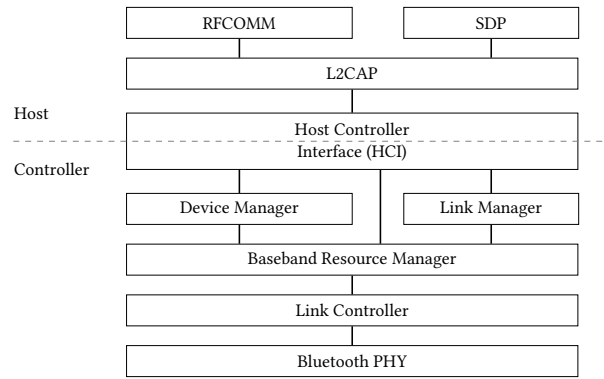
\begin{figure}
	\begin{center}
	\scalebox{0.75}{
	\begin{tikzpicture}[minimum height=0.55cm, node distance=0.9cm]
		\node[draw,  minimum width=4cm] (rfcomm) {RFCOMM};
		\node[draw, right=of rfcomm, minimum width=3.1cm] (sdp) {SDP};
		\node[draw, below=of rfcomm.west, minimum width=8cm, anchor=west] (l2cap) {L2CAP};
		\node[draw, below=of l2cap.west, minimum width=8cm, minimum height=1cm, anchor=west, 
				yshift=-0.3cm, align=center] (hci) {Host Controller \\Interface (HCI)};
		\node[draw, below=of hci.west, minimum width=4cm, anchor=west, yshift=-0.3cm] (devmgr) {Device Manager};
		\node[draw, right=of devmgr, minimum width=3.1cm] (linkmgr) {Link Manager};
		\node[draw, below=of devmgr.west, minimum width=8cm, anchor=west] (brm) {Baseband Resource Manager};
		\node[draw, below=of brm.west, minimum width=8cm, anchor=west] (linkctrl) {Link Controller};
		\node[draw, below=of linkctrl.west, minimum width=8cm, anchor=west] (phy) {Bluetooth PHY};
		\draw[thick] (rfcomm.south) -- (rfcomm.south |- l2cap.north);
		\draw[thick] (sdp.south) -- (sdp.south |- l2cap.north);
		\draw[thick] (l2cap.south) -- (hci.north);
		\draw[thick] (devmgr.north |- hci.south) -- (devmgr.north);
		\draw[thick] ([xshift=0.5cm] hci.south) -- ([xshift=0.5cm] brm.north);
		\draw[thick] (linkmgr.north |- hci.south) -- (linkmgr.north);
		\draw[thick] (devmgr.south) -- (devmgr.south |- brm.north);
		\draw[thick] (linkmgr.south) -- (linkmgr.south |- brm.north);
		\draw[thick] (brm.south) -- (linkctrl.north);
		\draw[thick] (linkctrl.south) -- (phy.north);

		\draw[dashed, color=gray] ([xshift=-2cm] hci.west) -- ([xshift=0.5cm] hci.east);
		\node[left=of hci.west, anchor=west, xshift=-1.2cm, yshift=0.5cm] (host) {Host};
		\node[left=of hci.west, anchor=west, xshift=-1.2cm, yshift=-0.5cm]  (controller) {Controller};
	\end{tikzpicture}
	}
	\end{center}
	\caption{Architecture of the Bluetooth Protocol Stack.}
	\label{fig:bluetooth_architecture}
\end{figure}

Unfortunately, compared to Wi-Fi there is no similarly easy way to monitor and manipulate the behavior of lower Bluetooth layers depicted in \autoref{fig:bluetooth_architecture}. On the one hand, professional equipment targeted
at hardware developers exists, but is very expensive.
On the other hand, modifiable open-source platforms still struggle with fundamental problems such as 
frequency-hopping characteristics of the Bluetooth \ac{PHY}.



This paper is the groundwork for a platform focused on Bluetooth low layer protocol modifications and security. It
targets the firmware of the \emph{Broadcom} Bluetooth chipset named \emph{BCM4339},
which resides inside the \emph{Nexus 5}. 
We tested and extended parts of the functionality to \emph{Nexus 6P}, \emph{Samsung Galaxy S6/S6 edge}, \emph{Raspberry Pi 3/3+} and a Bluetooth 5.0 IoT evaluation kit.

%
%


%

On top of reverse engineering results, the outcome of this work is
the research and analysis framework \mytool
 which enables direct interaction with
\emph{Broadcom} firmware internals at runtime. The framework consists of a flexible
\emph{Python} library which acts as an interface to the firmware over the \ac{ADB}.
It is supplemented with an interactive front-end which supports live analysis of the firmware and low-level
Bluetooth activities. The most important capabilities of \mytool are:
\begin{itemize}
	\item modifying arbitrary memory regions including \ac{ROM},
	\item run arbitrary code in the context of the running firmware,
	\item send arbitrary \ac{HCI} commands to the chip,
	\item establish connections to non-visible devices, and
	\item enable monitor mode and injection for the \ac{LMP}.
\end{itemize}
Access to \ac{LMP} stands out for transforming an off-the-shelf low-cost smartphone into an \ac{LMP} monitor and injection device, while enabling security testing.
To the best of our knowledge, there exists no openly available solution to monitor
and craft \ac{LMP} messages in the context of Bluetooth connections.
Such a tool is especially useful for researching and testing other Bluetooth
devices on the \ac{LM} layer, which controls important features such as security parameters and frequency settings.

To demonstrate the capabilities of \mytool, we use it to test for known Bluetooth bugs. On top of this, we detect a new severe vulnerability inside the firmware, which affected a huge fraction of \emph{Broadcom} Bluetooth chips in use in December 2018. We perform the following security demos and contributions with \mytool:
\begin{itemize}
	\item test for user interaction behavior of pairing devices without input and output capabilities~\cite{hypponen2007nino},
	\item test for a known \ac{ECDH} pairing vulnerability~\cite{2018:biham},
	\item \ac{MAC} address filter within \ac{LMP} to improve security,
	\item emulation framework to fuzz malicious payloads and trace their call graph, and
	\item discovery of a severe security vulnerability (CVE-2018-19860) that allows to crash various \emph{Broadcom} Bluetooth firmwares and execute a subset of functions within them.
\vspace{3em} 
\end{itemize}

\mytool is publicly available including various demos\footnote{\url{https://github.com/seemoo-lab/internalblue}}.
We disclosed all issues reported before handing in this paper to \emph{Broadcom}, 
who has acknowledged them and has already provided fixes to vendors.

This work is structured as follows.
\autoref{sec:reversing} describes relevant \emph{Broadcom} specifications and how to reverse engineer the firmware.
Based on this knowledge, the implementation of \mytool is explained in \autoref{sec:implementation}, including the application of an \ac{LMP} toolkit.
\mytool is then used to check for multiple known issues in the Bluetooth standard in \autoref{sec:old_exploits}.
Further investigation in \autoref{sec:new_exploits} leads to \emph{Broadcom} specific issues that allow code execution within Bluetooth firmware.
\autoref{sec:old_exploits} and \autoref{sec:new_exploits} contain the main contributions of this paper---readers who are not interested in the efforts required to open a closed source firmware and do not want to reproduce results can skip \autoref{sec:reversing} and \autoref{sec:implementation}.
Related work is listed in \autoref{sec:related}.
Results are discussed in \autoref{sec:discussion}. 
\autoref{sec:conclusion} concludes this paper.

%
%
%

\section{Firmware Reverse Engineering}
\label{sec:reversing}

%
%

%



This section summarizes the information on the internal structure and
functioning of the \emph{Broadcom} \emph{BCM4339} Bluetooth controller. 
The information has been gathered through reverse engineering the controller's firmware according to the Bluetooth specifications~\cite{2016:SIG}, and also from the datasheet in~\cite{2016:cypress}. 
Some functions can be clearly mapped to standardized Bluetooth procedures, because they use uniquely specified values.
However, going from machine code to assembly and data sections and then naming functions is a tedious process and nobody publicly did this for \emph{Broadcom} Bluetooth chips before.  The normal procedure for experimenting with lower layer Bluetooth would be signing a non-disclosure agreement with a vendor. However, in that case, results would not be openly publishable to the community.
Reverse engineering techniques presented in this section apply to all \emph{Broadcom} Bluetooth chips and might also be helpful in similar projects.

Readers who do not want to add their own functions to \mytool but at least use already existing functions themselves can directly proceed to \autoref{sec:implementation}. Readers who are only interested in applications and outcomes can proceed to \autoref{sec:old_exploits}.

\newpage 

\subsection{Hardware Architecture}

\begin{table}[b]
	\caption{Memory Map of the Firmware's Address Space.}
	\label{tab:memory_map}
	\small
	\begin{tabular*}{\linewidth}{ @{\extracolsep{\fill}} l l r l l}
	
		Start & End & Size & Type & Description \\
		\hline		
		\hex{000000}   & \hex{090000} & 576 kB & ROM & Firmware\\
		\hex{0D0000}   & \hex{0D8000} &  32 kB & RAM & \emph{Patchram}\\
		\hex{200000}   & \hex{228000} & 160 kB & RAM & Stack \& Heap\\
		\hex{260000}   & \hex{268000} &  32 kB & ROM & Data \& Jumptables\\
		\hex{310000}   & \hex{310400} &   1 kB &  IO & \emph{Patchram} Control\\
		
	\end{tabular*}
\end{table}

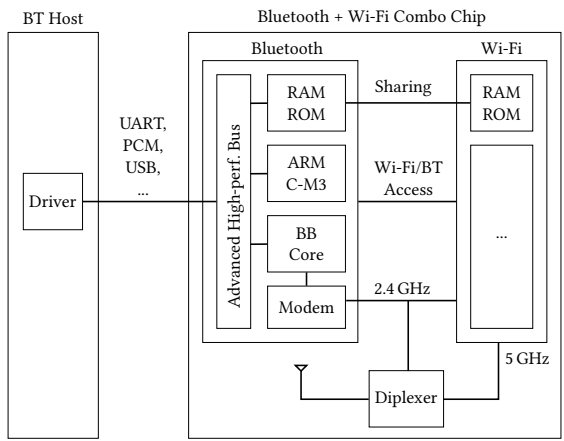
\begin{figure}[b]
	\begin{center}
	\scalebox{0.75}{
	\begin{tikzpicture}[anchor=south west]
		\node[draw, minimum height=7.2cm, minimum width=1.6cm] (host) at (1.8, 0.3){};
		\node[draw, minimum height=1cm, minimum width=1cm, align=center] (driver) at (2.07, 4) {Driver};
		\node[draw, minimum height=7.2cm, minimum width=6.6cm] (chip) at (5, 0.3) {};
		\node[draw, minimum height=5cm, minimum width=2.75cm, align=center] (btstack) at (5.25, 2) {};
		\node[draw, minimum height=4.5cm, minimum width=0.6cm, align=center] (ahb) at (5.5, 2.25) {};
		\node[rotate=90] at (6.1, 2.5) {Advanced High-perf. Bus};
		\node[draw, minimum height=1cm, minimum width=1.38cm, align=center] (rambt) at (6.4, 5.75) {RAM\\ ROM};
		\node[draw, minimum height=1cm, minimum width=1.38cm, align=center] (otherbt) at (6.4, 4.5) {ARM\\ C-M3};
		\node[draw, minimum height=1cm, minimum width=1.38cm, align=center] (bbc) at (6.4, 3.25) {BB\\ Core};
		\node[draw, minimum height=0.75cm, minimum width=1.38cm, align=center] (modem) at (6.4, 2.25) {Modem};
		\node[draw, minimum height=1cm, minimum width=1cm, align=center] (diplexer) at (8.2, 0.5) {Diplexer};
		\node[draw, minimum height=1cm, minimum width=1.1cm, align=center] (ramWi-Fi) at (10, 5.75) {RAM\\ ROM};
		\node[draw, minimum height=3.25cm, minimum width=1.1cm, align=center] (otherWi-Fi) at (10, 2.25) {...};
		\node[draw, minimum height=5cm, minimum width=1.6cm, align=center] (Wi-Fistack) at (9.75, 2) {};
		\node[minimum width=1.6cm] (hostlabel) at (1.8,7.54) {BT Host};
		\node[minimum width=6cm] (chiplabel) at (5.25,7.5) {Bluetooth + Wi-Fi Combo Chip};
		\node[minimum width=2cm] (chiplabel) at (5.75,7) {Bluetooth};
		\node[minimum width=1.6cm] (chiplabel) at (9.75,7) {Wi-Fi};
		\draw[-,draw,thick] (driver) -- node[xshift=-0.65cm,align=center] {UART, \\ PCM, \\ USB, \\ ...} (ahb);
		\draw[-,draw,thick] (rambt) -- node[xshift=-0.7cm,align=center] {Sharing} (ramWi-Fi);
		\draw[-,draw,thick] (btstack) -- node[xshift=-0.7cm,align=center] {Wi-Fi/BT \\ Access} (Wi-Fistack);
		\draw[-,draw,thick] (9.52, 1) -- (10.5,1) -- node[align=center] {5\,GHz} (10.5,2);
		\draw[-,draw,thick] (7.8,2.75) -- node[xshift=-0.6cm,align=center] {2.4\,GHz} (9.75,2.75);
		\path[-,draw,thick] (8.9,2.75) -- (8.9,1.5);
		
		\path[-,draw,thick] (8.2,1) -- (7,1) -- (7,1.5) -- (6.9,1.6) -- (7.1,1.6) -- (7,1.5);
				
		\path[-,draw,thick] (otherbt) -- (6.1,5);
		\path[-,draw,thick] (rambt) -- (6.1,6.25);
		\path[-,draw,thick] (bbc) -- (6.1,3.75);
		\path[-,draw,thick] (bbc) -- (modem);
		
	\end{tikzpicture}
	}
	\end{center}
	\caption{Simplified Combo Chip Architecture~\cite{2016:cypress}.}
	\label{fig:combo}
\end{figure}

The \emph{BCM4339} combines a Wi-Fi IEEE 802.11ac and Bluetooth 4.1
transceiver on a single chip. A simplified architecture overview is depicted in \autoref{fig:combo}. Inside the chip, Wi-Fi and
Bluetooth units operate on separated processor cores.
The Bluetooth processor is an \emph{ARM Cortex-M3}, which has \SI{196}{\kilo\byte} \ac{RAM} and
\SI{608}{\kilo\byte} \ac{ROM}. Both, \ac{RAM} and \ac{ROM},
are split up into two sections respectively as seen in \autoref{tab:memory_map}.
In addition, the chip contains a Bluetooth modem controlled by the
\emph{Bluetooth Baseband Core}.
This component handles the lower layers of the Bluetooth protocol stack, while
the higher layers such as the link and device manager are implemented in the
\emph{Cortex-M3}. Components are interconnected by an \emph{ARM} \ac{AHB}.

Communication with the host system is done via \ac{HCI} over a
\ac{UART} interface or \ac{USB}.
Host communication includes a \ac{PCM} connector for transferring audio
data. 

\subsection{Reading, Writing, and Executing RAM}

\ac{HCI} provides a uniform method for
the host to access controller capabilities and send or receive
data from underlying Bluetooth links. 
The protocol is based on commands the host sends to the controller, and events the controller sends to the host.
 Accessing \ac{HCI} on the host
requires superuser rights, the concrete implementation depends on 
the operating system.
\ac{HCI} is never directly sent over the air.

Apart from the standardized \ac{HCI} commands~\cite{2016:SIG},
the \emph{BCM4339} implements many vendor specific commands partially listed in
the datasheet~\cite{2016:cypress2}. The \mbox{\lstinline{Write_RAM},}
\lstinline{Read_RAM}, \lstinline{Download_Minidriver} and
\lstinline{Launch_RAM} commands are relevant in order to understand
firmware patching and debugging, as described in \autoref{sec:architecture}. 

\lstinline{Read_RAM}
reads memory chunks from the chip's
\ac{RAM} and \ac{ROM} address space and sends them back to the host in the \ac{HCI} command complete event. 
Reading
unmapped addresses causes the chip to crash.

\lstinline{Write_RAM} can be used to
write memory chunks to \ac{RAM} regions of the chip's memory. If 
attempting to write to a \ac{ROM} section, the command has no
effect.

\lstinline{Download_Minidriver}
will put the device into a special mode in which it is safe to receive patches, normal Bluetooth activity is disabled. Only a very reduced subset of \ac{HCI}
commands is interpreted in  download mode, including \lstinline{Read_RAM}, 
\lstinline{Write_RAM} and \mbox{\lstinline{Launch_RAM}.} \lstinline{Launch_RAM} is being used 
to exit download mode and reboot into the normal Bluetooth firmware.
Any downloaded \ac{ROM} patches are applied at an early stage during this reboot.

\lstinline{Launch_RAM} continues code execution on the chip at a
specified address using \emph{Thumb} mode. One of the intended use cases is to exit \lstinline{Download_Minidriver} mode by passing the pseudo-address \hex{FFFFFFFF} to \lstinline{Launch_RAM}. This  applies the downloaded
patches, quits download \lstinline{Download_Minidriver}, and goes back into Bluetooth mode. Another  use case is to jump to the entry point of a so-called \emph{Minidriver}
which was previously loaded into \ac{RAM} using the \lstinline{Download_Minidriver}
and \lstinline{Write_RAM} commands. \\

\noindent Support of \lstinline{Write_RAM}, \lstinline{Read_RAM}, and \lstinline{Launch_RAM} is essential for \mytool. These \ac{HCI} commands can be considered security critical, as the installed patches
are not checked for a signature or similar from the vendor; hence, allowing malicious code installation. They exist throughout all chip versions---including one of the most recent chips \emph{CYW20735}. Technically, this means \mytool can be ported to all current \emph{Broadcom} and \emph{Cypress} Bluetooth chips.

\subsection{Updating Firmware and ROM}
\label{sec:firmware_update}

The \emph{Android} Bluetooth driver includes an HCD file with firmware updates and patches in its initialization
procedure. By first reading the name of the chip via an \ac{HCI} command it is
able to search for a matching HCD file inside the \lstinline{/system/vendor/firmware}
directory. It applies the HCD file and then continues with normal chip initialization. 

An HCD file contains
multiple \lstinline{Write_RAM} commands adding new functions
and data to the \ac{RAM} sections of the chip. 
The HCD file terminates with a \lstinline{Launch_RAM} command with \hex{FFFFFFFF} as argument
in order to reboot into Bluetooth mode and apply the patches.


Patches to \ac{ROM} cannot be applied with \mbox{\lstinline{Write_RAM},} but with a special \emph{Patchram} mechanism.
Although this mechanism
is mentioned in the \emph{BCM4339} datasheet~\cite{2016:cypress}, it 
is not publicly documented and therefore had to be reverse engineered. 
All \ac{ROM}
patches are collected in a list and written to a custom 
\ac{RAM} address. The list is formatted as a chain of \ac{TLV} objects. Each entry starts with a one-byte type field
followed by a two-byte length of the payload. 
Finally, \lstinline{Launch_RAM} will exit the
\lstinline{Download_Minidriver} state, apply patches to ROM and
enter normal Bluetooth mode.

The \ac{TLV} type \hex{08} is used for single \SI{32}{bit} patches to the ROM (called \emph{Patchram}). The
length of the value is fixed to 15 bytes with the following semantic:

\vspace{0.4em}
\begin{tabular}{c c c c c}
	1 Byte & 4 Bytes & 4 Bytes & 2 Bytes & 4 Bytes\\
	\hline
	{\lstinline!#slot!} & {\lstinline!address!} & {\lstinline!value!} & \hex{0000} & \emph{unknown}
\end{tabular}
\vspace{0.4em}

The final effect of this \emph{Patchram} object is: the new \lstinline{value} virtually overwrites four bytes of memory inside the \ac{ROM}, which are located at the target \lstinline{address}.
\emph{Patchram} patches only last until the next reset.
Internally this is done by
writing the \lstinline{value} into a table at address \hex{D0000}
(\ac{RAM}) and the target \lstinline{address} into another table at address
\hex{310000} (hardware register). For both tables the index is given by the
\lstinline{slot} number and the tables are capable of holding up to 128 entries
which correspond to slot numbers 0 to 127.

Overwriting 128 values of 4 bytes in \ac{ROM} is not much, but this is sufficient to inject branch commands into the beginnings of functions and then redefining them inside the RAM. The \emph{Nexus 5} HCD file in \emph{Android 6} already uses 113 \emph{Patchram} slots---therefore not only limiting the amount of modifications \mytool can install but also future vendor bugfixes. \emph{Broadcom} has realized that this issue reduces patching abilities, and thus provides more \emph{Patchram} slots on newer chips, for instance, the \emph{Nexus 6P} has now 192 slots. Yet, newer chips are also shipped with more code, raising the probability of erroneous code that requires fixes.

To overcome this shortage of \emph{Patchram} slots, meta patches being responsible for multiple patches could be inserted in popular parts of the program flow. However, such meta patches require more code and free \ac{RAM} is also a very limited resource.

%

\subsection{Link Manager Functions}
\label{ssec:link_manager}

The \ac{LM} communicates with the \acp{LM} of other controllers via the
\ac{LMP} in order to establish and control links between devices. Its tasks
include to set-up and control logical transports and links,
synchronize clocks between Bluetooth master and slaves,
control frequency hopping sequence, and set-up
authentication and encryption. Relevant information exchanged on the \ac{LM} is
then passed by the firmware via \ac{HCI} to the host.

At the core of the \ac{LM} in the \emph{BCM4339} firmware is a handler table
for each \ac{LMP} \lstinline{opcode}. The table starts at the address \hex{261610} in the
second half of the \ac{ROM}. Each table entry has a length of 8 bytes and
contains the following values:

\vspace{0.4em}
\begin{tabular}{c c c c c}
	4 Bytes & 1 Byte & 3 Bytes\\
	\hline
	{\lstinline!*function!} & {\lstinline!lmp_packet_len!} &  \emph{additional info}
\end{tabular}
\vspace{0.4em}

A thread periodically checks for incoming \ac{LMP} packets
and calls the function \lstinline{LMP_dispatcher} if new packets arrive.
A global data structure \lstinline{lm_curCmd} at address \hex{200478} holds information about the
received packet currently being processed. The \ac{LMP} dispatcher reads the \lstinline{opcode} from \lstinline{lm_curCmd},
retrieves the corresponding entry from the \ac{LMP} function table and
finally calls the handler function from the retrieved entry.

Handler functions take a pointer to a structure representing the respective
connection as their only argument. They get the payload for the received packet
from \lstinline{lm_curCmd}. If the received packet requires
a response, the handler crafts a new \ac{LMP} packet and calls
\lstinline{send_LMP_packet}. This function puts the packet into a transmit buffer
for the respective connection.

\subsection{Reversing Evaluation Kits}
Evaluation boards such as \emph{CYW920735Q60EVB-01} for the \emph{CYW20735} Bluetooth 5.0 chipset are supported by \emph{WICED Studio}~\cite{cypress_cyw20735}. They do not only run a basic Bluetooth firmware---which has only minor changes compared to other versions---but also an application like a temperature sensor providing readings. Simple applications not requiring their own processor make this type of chips suitable for IoT implementations. \emph{WICED} applications are tightly coupled to firmware, for example they use the same thread initialization function. Investigating similarities we found a binary file \lstinline{patch.elf} which defines almost all symbols, a total of 11813 function and 2818 object names. Note that access to firmware definitions usually requires a non-disclosure agreement with the vendor, and the binary format of \lstinline{patch.elf} abstracts them in a way that is hard to search for. When finding this source of information, we already had invested months of reverse engineering \emph{Nexus 5} firmware---all \ac{LM} functions were located by searching for opcodes defined in the standard. 

Locating functions in \emph{CYW20735} also helps to reverse engineer other \emph{Broadcom} chips. Binary diffing tools can apply heuristics to locate similar functions. Yet, when using \emph{IDA Pro BinDiff}~\cite{bindiff} only 6\% of the named objects and functions are found in the \emph{Nexus 5} firmware, more than half of them being false positives, despite only selecting matches with at least 90\% similarity. The good news is that a non-automated search shows way more correct similarities including complete libraries that did not change over 6 years of development. Low variation in the code and missing obfuscation due to performance and space requirements makes porting \mytool firmware patches comparably easy.
However, there are also new features introduced over those 6 years, updated libraries and architectural changes, as well as variances in compilation that make \emph{BinDiff} giving bad results. Thus, previous reversing of the \emph{Nexus 5} smartphone still provides interesting insights.

Despite knowing many original function names from \emph{WICED Studio}, we continue using our own comprehensible names throughout this paper. For the curious reverser, the most relevant functions \lstinline{LMP_dispatcher} and \lstinline{send_LMP_packet} are originally called \lstinline{lm_HandleLmpReceivedPdu} and \lstinline{DHM_LMPTx}.


\begin{table}[b]
	\caption{Currently Supported Chips by \mytool, \\ * refers to the diagnostic protocol~\cite{wisec_diag}.}
	\label{tab:chips}
	\small
	\begin{tabularx}{\linewidth}{ @{\extracolsep{\fill}} p{0.9cm} c c c p{3cm}}
	
		Chip & 				Patchram &	LMP  & Tracepoint & Device \\
		\hline		
		\emph{BCM4339} & 	\checkmark 	& 	\checkmark &	\checkmark &		\emph{Nexus 5 / Xperia Z3} \\	
		\emph{BCM4358} & 	\checkmark 	& 	\checkmark &	 $\times$ &	\emph{Nexus 6P / \newline Samsung Galaxy S6 [edge]}  \\
		\emph{BCM4343} & 	\checkmark 	& 	* &	$\times$ &		\emph{Raspberry Pi 3} \\
		\emph{BCM4345} & 	\checkmark 	& 	* & $\times$ &			\emph{Raspberry Pi 3+} \\
		\emph{CYW20735} & 	\checkmark 	& 	* &	\checkmark &			\emph{Evaluation Kit}  \\

	\end{tabularx}
\end{table}

\section{Basic InternalBlue Functions}
\label{sec:implementation}

Knowledge gathered in the previous section is used to create \mytool, an
open and extendable Bluetooth research framework based on \emph{BCM4339}
and other \emph{Broadcom} chips. The aim is to repurpose off-the-shelf consumer devices as analysis tools for experimenting with the lower
layers of Bluetooth connections. This is done by patching the controller's firmware and writing additional software for the host side which communicates with
the firmware over \ac{HCI}.

The first basic firmware tweaks focus on integrating \ac{LMP} functionality into \mytool. 
As shown in \autoref{fig:bluetooth_architecture}, \ac{LMP} resides below \ac{HCI}, with the latter already being accessible on \emph{Android}. Monitoring and modification of layers below allows for more fine-grained observations and changes, and details on \ac{LMP} inside the firmware give insights where to integrate own functions. Therefore, we implement an \ac{LMP} monitor and then extend it with injection capabilities.

An overview of currently supported chips and tested features is given in \autoref{tab:chips}. Reading and writing \ac{RAM} worked on any \emph{Broadcom} chip we encountered so far; thus, we only list the special \emph{Patchram}, \ac{LMP} and tracepoint debugging features. Patches for those features depend solely on the chip, but we list the device used for testing. An alternate way to implement \ac{LMP} without patches is a proprietary diagnostic protocol~\cite{wisec_diag}.
 Check out the current repositories to get an up to date list.

\subsection{Architecture}
\label{sec:architecture}

The architecture of the framework allows for close inspection and
manipulation of the Bluetooth controller while being actively used by
\emph{Android}. \autoref{fig:system_overview} shows an overview of the system
which consists of an \emph{Android} device with an embedded \emph{Broadcom} Bluetooth
controller, and a Linux host running the \mytool framework, which is written in \emph{Python}.
The \emph{Android} device is connected to the host via \ac{ADB}.

The framework relies on debug functionality in the \emph{Android} Bluetooth
stack such as the \emph{HCI Snoop Log} feature.  This feature writes a capture
file of all \ac{HCI} packets which were transferred between the \emph{Android}
host and the Bluetooth controller. Despite \ac{LMP} resides below \ac{HCI}, adjusted \ac{LMP} traffic can be forwarded over this interface to the host.  Once \emph{HCI Snoop Log} has been enabled through
the \emph{Android} developer settings, a capture file is written to the
internal memory of the \emph{Android} device. It uses the \emph{Snoop Packet
Capture} format which is specified in RFC 1761~\cite{RFC1761}. Each \ac{HCI} packet is
encapsulated in a \emph{Snoop Record} which also stores length and time
information in a \emph{Record Header}.

If the \emph{Android} Bluetooth stack is compiled with debugging features enabled (\lstinline{BT_NET_DEBUG}), networking features in \lstinline{aosp/platform/system/bt/+/master/hci/src/btsnoop_net.c} are activated and \emph{Android} opens two TCP ports listening on localhost~\cite{android_bt2}: 
\begin{description}
	\item[127.0.0.1:8872] outputs every \ac{HCI} packet exchanged between
		the \emph{Android} Bluetooth stack and the Bluetooth controller using the
		\emph{Snoop} file format.
	\item[127.0.0.1:8873] accepts \ac{HCI} commands in the format of \emph{Snoop Records},
		which the \emph{Android} Bluetooth stack will send the Bluetooth controller.
\end{description}
Compilation with \lstinline{BT_NET_DEBUG} enabled was tested and confirmed to be working  on \ac{AOSP} \emph{Android} 6.0.1 and 7.1.2, and \emph{LineageOS} 14.1.

If the host system in which the Bluetooth controller resides is not \emph{Android} but \emph{Linux} with \emph{BlueZ}, sending \ac{HCI} commands works out of the box without further changes using \lstinline{AF_BLUETOOTH} sockets.

\begin{figure}[b]
	\centering
	\vspace{3.8px}
	\includegraphics[width=\columnwidth]{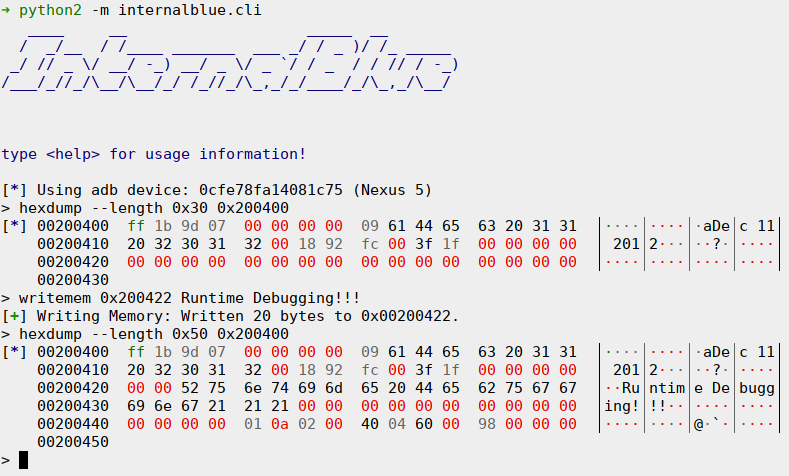}
	\caption{\mytool CLI.}
	\label{fig:screenshot_cli}
\end{figure}

\begin{figure}[b]
	\begin{center}
	\scalebox{0.75}{
	\begin{tikzpicture}[anchor=south west]
		\node[draw, minimum height=7cm, minimum width=4cm] (host) at (0, 0.5){};
		\node[draw, minimum height=1.5cm, minimum width=3cm, align=center] (core) at (0.5, 4.5) {Core\\Framework};
		\node[draw, minimum height=1cm, minimum width=3cm, align=center] (cli) at (0.5, 6) {CLI / Script};
		\node[draw, minimum height=7cm, minimum width=4cm] (android) at (7, 0.5) {};
		\node[draw, minimum height=.7cm, minimum width=3cm] (bcm) at (7.5, 0.9) {\emph{BCM4339}};
		\node[draw, minimum height=4.5cm, minimum width=3cm, align=center] (btstack) at (7.5, 2.5) {Android\\Bluetooth\\Stack\\(compiled with\\debugging\\enabled)\\\\\\\\};
		\node[draw, circle, minimum height=0.2cm, minimum width=0.2cm] (port1) at (8, 3.4) {};
		\node[draw, circle, minimum height=0.2cm, minimum width=0.2cm] (port2) at (8, 3) {};
		\node[draw, circle, minimum height=0.2cm, minimum width=0.2cm] (port1host) at (3, 3.4) {};
		\node[draw, circle, minimum height=0.2cm, minimum width=0.2cm] (port2host) at (3, 3) {};

		\node[minimum width=4cm, minimum height=2em] (hostlabel) at (0,7.5) {Host System};
		\node[minimum width=4cm, minimum height=2em] (androidlabel) at (7,7.54) {Android Device};
		\node[minimum width=4cm] (internalbluelabel) at (0,7) {\mytool};
		\node (port1label) at (7.8,3.6) {\footnotesize{8872}};
		\node (port2label) at (7.8,2.6) {\footnotesize{8873}};
		\path[->,draw,thick] (core.east) -- node [xshift=-1.6cm, text width=3cm, align=center] {ADB \\ (USB or Wi-Fi)} (btstack.west |- core.east);
		\path[<-,draw,dashed,thick] (port1host) -- (port1);
		\path[->,draw,dashed,thick] (port2host) -- node [xshift=-1.3cm,yshift=-1.3cm,align=center]
				{ADB Port\\Forwarding} (port2);
		\draw[<-,draw,thick] (port1.east) .. controls (9.03,3.4) .. (bcm);
		\draw[->,draw,thick] (port2.east) .. controls (9.03,3) .. (bcm);
		\draw[<->,draw,thick] (9,4) -- node[yshift=-1.2cm,align=center] {HCI over\\UART} (bcm);
		\draw[->,draw,thick] (port1host.west) .. controls ([xshift=0.1cm]core.south |- port1host.west) .. ([xshift=0.1cm]core.south);
		\draw[<-,draw,thick] (port2host.west) .. controls ([xshift=-0.1cm]core.south |- port2host.west) .. ([xshift=-0.1cm]core.south);
	\end{tikzpicture}
	}
	\end{center}
	\caption{System Overview of \mytool.}
	\label{fig:system_overview}
\end{figure}

\subsection{Command Line Interface}

The \mytool \ac{CLI} offers a multicolored \ac{UI} shown
in \autoref{fig:screenshot_cli}. It implements various commands to
conveniently access the firmware on any \emph{Broadcom} or \emph{Cypress} Bluetooth chip:
\begin{itemize}
	\item read and present memory in various different representations including disassembly,
	\item write data to memory in different representations,
	\item search for data in the memory of the firmware,
	\item execute custom code snippets in the firmware,
	\item activate monitor mode for \ac{HCI} and forward
		the received packets to \emph{Wireshark}~\cite{wireshark},
	\item inject arbitrary \ac{HCI} packets, and
	\item establish connections by \ac{MAC} address.
\end{itemize}

Moreover, the following extensions are available as assembly patches for \emph{BCM4339} on the \emph{Nexus 5}:
\begin{itemize}
	\item activate monitor mode for \ac{LMP} over \ac{HCI},
	\item inject arbitrary---also invalid---\ac{LMP} packets, and
	\item set tracepoints.
\end{itemize}

Executed commands are saved in a history file and it is possible to browse to an earlier
command by pressing the arrow-up key as it is known from similar \acp{CLI}.

\ac{HCI} and \ac{LMP} monitoring modes directly open \emph{Wireshark} with correct connection parameters, while the latter also applies the required patches to the firmware. To use \mytool simply as an \ac{HCI} live monitor and injector on any \emph{Android} device---no matter if the device has a \emph{Broadcom} chipset or not---it is sufficient to recompile the \emph{Android} Bluetooth stack with debug features. To the best of our knowledge, \mytool is the only project providing \emph{Python} bindings to access \ac{HCI} on remote \emph{Android} devices.


%


\subsection{LMP Monitor and Injection Mode}
\label{ssec:lmp}

The \ac{LM} is responsible
for pairing, managing connections, as well as many other interesting processes which can now be monitored
and modified with the feature developed in this section.
In contrast to \ac{HCI}, its contents are sent over the air between devices.
As \ac{LMP} operates exclusively beneath the \ac{HCI} protocol layer, it
is usually impossible to monitor or directly manipulate it from the host.
By using the knowledge gained in \autoref{ssec:link_manager} and the
patching capabilities of \mytool, \emph{BCM4339} firmware
can be modified to relay \ac{LMP} packets to the host and provide an interface
for injecting arbitrary \ac{LMP} packets into an established connection.

\begin{figure}
	\centering
	\includegraphics[width=\columnwidth]{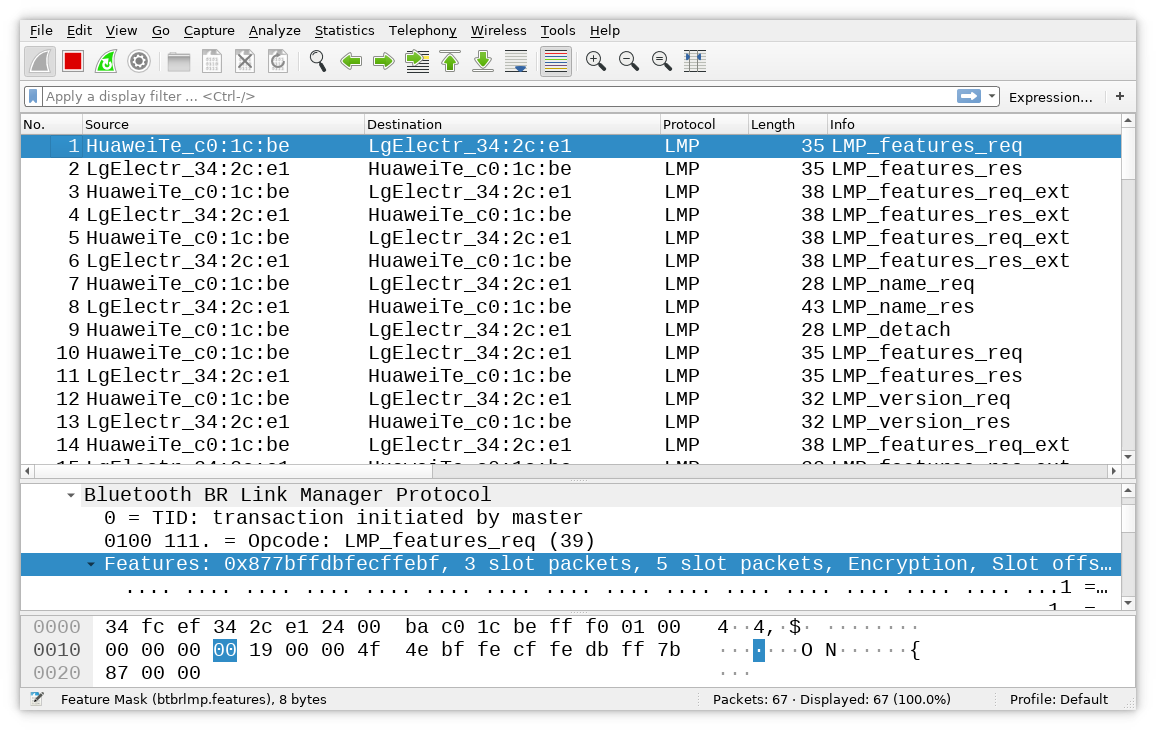}
	\caption{LMP Monitor.}
	\label{fig:lmp_monitor}
\end{figure}

\begin{figure}
	\begin{center}
	\scalebox{0.75}{
	\begin{tikzpicture}[node distance=0.45cm]
		\node[draw, minimum height=1.5cm, minimum width=3cm,align=left] (dispatcher) {\lstinline{LMP_dispatcher}\\function\\\\};
		\node[draw, minimum height=1.5cm, minimum width=3cm,align=left, below=of dispatcher] (sendlmp) {\lstinline{send_LMP_packet}\\function\\\\};
		\node[draw, minimum height=0.7cm, minimum width=1.5cm, right=of dispatcher, fill=white, xshift=-1.5cm] (hook1) {HOOK};
		\node[draw, minimum height=0.7cm, minimum width=1.5cm, right=of sendlmp, fill=white, xshift=-1.5cm] (hook2) {HOOK};
		\node[draw, minimum height=1cm, minimum width=1.5cm, rounded corners=0.3cm, right=of hook1, fill=white] (hci) {HCI};
		\node[draw, diamond, aspect=2.5, right=of hci, align=center] (decision) {LMP monitor event?};
		\node[draw, minimum height=1cm, minimum width=3cm,align=center, below=of decision] (pcap) {encapsulate\\into \emph{pcap}};
		\node[draw, minimum height=0.75cm, minimum width=3cm,align=center, below=of pcap] (wireshark) {\emph{Wireshark}};
		\path[->,draw,thick] (hook1.east) -- (hci.west);
		\path[->,draw,thick] (hook2.east) -| ([xshift=-0.2cm]hci.west) -- (hci.west);
		\path[->,draw,thick] (hci.east) -- (decision.west);
		\path[->,draw,thick] (decision.south) -- node[xshift=0.5cm] {YES} (pcap.north);
		\path[->,draw,thick] (pcap.south) -- (wireshark.north);
		\path[->,draw,thick,dashed] (decision.east) -- node[yshift=0.2cm] {NO} ([xshift=0.5cm]decision.east);
		\draw[dashed, color=gray] ([yshift=-3cm] hci.south) -- (hci.south);
		\draw[dashed, color=gray] (hci.north) -- ([yshift=1.25cm] hci.north);
		\node[above=of hci, anchor=west, yshift=0.5cm, xshift=0.2cm] (host) {Host};
		\node[above=of hci, anchor=east, yshift=0.5cm, xshift=-0.2cm] (controller) {Controller};
	\end{tikzpicture}
	}
	\end{center}
	\caption{Receive Path of the LMP Monitor Mode.}
	\label{fig:lmp_receive_path}
\end{figure}
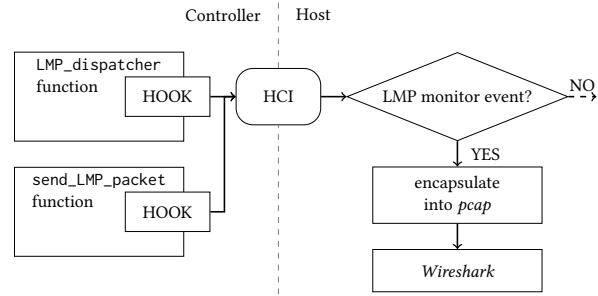

The receive path of the monitor mode patch is illustrated in \autoref{fig:lmp_receive_path}.
It starts at the \lstinline{LMP_dispatcher} function
that processes incoming packets by invoking the responsible handler
function for the specific \ac{LMP} \lstinline{opcode}. By hooking the dispatcher
function, the received packet can be copied and transferred to the host via a
custom \ac{HCI} event packet. In addition, the function \lstinline{send_LMP_packet} responsible for
sending \ac{LMP} packets to a remote device is also hooked in order to have
access to those packets on the host. At the host side, the monitor feature
installs an \ac{HCI} receiver callback, which listens for the custom \ac{HCI}
events, processes them and passes the \ac{LMP} packets to a \emph{Wireshark} instance. 
As \emph{Wireshark} is not able to dissect \ac{LMP} packets by default, it is
necessary to install a custom dissector plugin for this protocol. Fortunately,
the \emph{Ubertooth} project \cite{ubertooth:project} includes such a plugin that,
after minor modifications and functional bugfixes, can be used for this purpose.
The \mytool command \lstinline{monitor lmp start} automatically installs hooks for monitoring \ac{LMP} traffic inside the firmware and starts a \emph{Wireshark} instance as seen in \autoref{fig:lmp_monitor}.

Injection of arbitrary \ac{LMP} packets into an established Bluetooth
connection works by invoking the \lstinline{send_LMP_packet} function inside
the firmware, which in turn will enqueue the packet into the queue for outgoing
\ac{LMP} packets. To this end, the payload and a small assembly code stub are
written to the \ac{RAM} and invoked with the \lstinline{Launch_RAM} \ac{HCI}
command. The stub copies the payload into a freshly allocated
buffer and calls the \lstinline{send_LMP_packet} function with a pointer to
the corresponding connection structure. It also makes sure that the \ac{TID} bit
in the \ac{LMP} header is set correctly based on whether the device is
master or slave in the connection.
If the fuzzing feature is enabled over the command line, an additional patch skips opcode and packet length checks.

%

\section{Testing for Known Security Issues}
\label{sec:old_exploits}

%

Inside the \mytool \ac{CLI} we implement the feature to connect to any device by its \ac{MAC} address even without prior pairing, as described in \autoref{ssec:connect}.
As a demonstration of security research, we implement device testing for the so-called \emph{Niño} attack---pairing devices without input and output---in \autoref{ssec:nino} as well as an attack on \ac{ECDH} device pairing key exchange in \autoref{ssec:ecdh}.

Implementations of the \ac{ECDH} and \emph{Niño} tests are not contained as \mytool commands but as example files\footnote{\url{https://github.com/seemoo-lab/internalblue/tree/master/examples}} that use functions of \mytool as library. 
Since the connecting smartphone itself performs the attacks and there is no such thing as \ac{MITM}, then these implementations are only suitable for testing how other Bluetooth stacks react to \ac{ECDH} and \emph{Niño} attacks---we do not provide active attacks on a victim's established connections.

\subsection{Establishing Connections to ``Invisible'' Devices}
\label{ssec:connect}
Bluetooth devices can be invisible to scanning but still accept connections. This is the default behavior for a setup with a smartphone and a headset, with the headset connecting to the smartphone when turned on knowing its \ac{MAC} address from previous communication. Surprisingly to most users, initiating a connection request to a device does not require previous pairing. Everything that needs to be known is a valid \ac{MAC} address~\cite{2009:ossmann}. \mytool can list the array of all current connections and initiate new connections with the command syntax \lstinline{connect de:ad:be:ef:00:00}.

Existence of a connection inside the firmware's list is required to send \ac{LMP} packets. This list differs from the list of known devices on the host, it is just meant to manage connection statuses of ongoing connections. A connection is not required to be established to appear in this list, it is sufficient to initiate a connection. If a connection establishment is not successful, it will stay half a minute inside the firmware's list until it is deleted. 
This allows to send arbitrary \ac{LMP} commands to arbitrary devices without successful pairing.

\subsection{No Input No Output}
\label{ssec:nino}

%

Bluetooth provides pairing based on \ac{ECDH} key exchange with \ac{SSP} to be secure against passive and active \ac{MITM} attacks. 
Security against active \ac{MITM} is provided by visual number comparison or entering a number shown on one device with a keyboard on the other device. This protection requires both devices to have \ac{IO} capabilities, which is often not the case for \ac{IoT} gadgets and headsets. The fallback option in case of missing \ac{IO} capabilities is to perform key exchange without active \ac{MITM} protection, the so-called ``Just Works'' mode. The Bluetooth 5.0 standard explicitly marks this as being insecure, but does not suggest to show any warnings to the user.

Exchange of \ac{IO} capabilities supported by the devices takes place before pairing. Since there is no established root of trust between them at this stage, \ac{IO} capabilities are prone to active  attacks themselves---an active \ac{MITM} can replace capabilities by no input no output, also known as the \emph{Niño} attack~\cite{hypponen2007nino}.

We patch the \ac{SSP} state machine and claim our \emph{Nexus 5} has no display and no keyboard.
Our question is if vendors accept a ``Just Works'' pairing with asking a ``yes/no'' message or warn the user to check if the device they are pairing to really has no \ac{IO} capabilities. Even though certificate warnings in browser have been there for years to warn users, warning from potentially insecure Bluetooth pairings seems to be uncommon. Neither \emph{Android} nor \emph{iOS} warn users about this. 

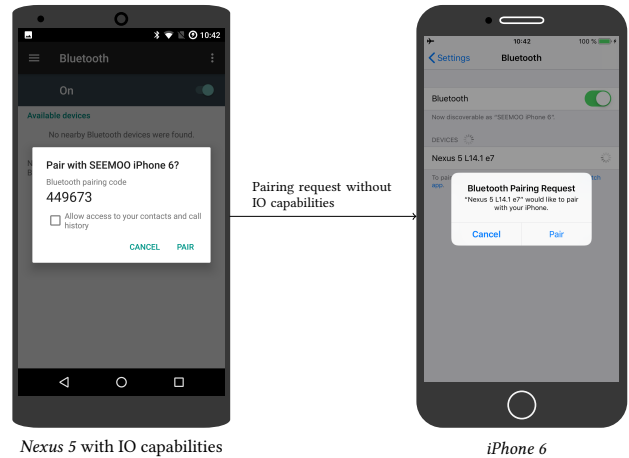
\begin{figure}
	\begin{center}
	\hspace{-0.23cm}\scalebox{0.75}{
    \begin{tikzpicture}[minimum height=0.55cm, node distance=0.7cm]

        \node[inner sep=0pt] (iphone) at (7.1,0)
    {\includegraphics[height=7.5cm]{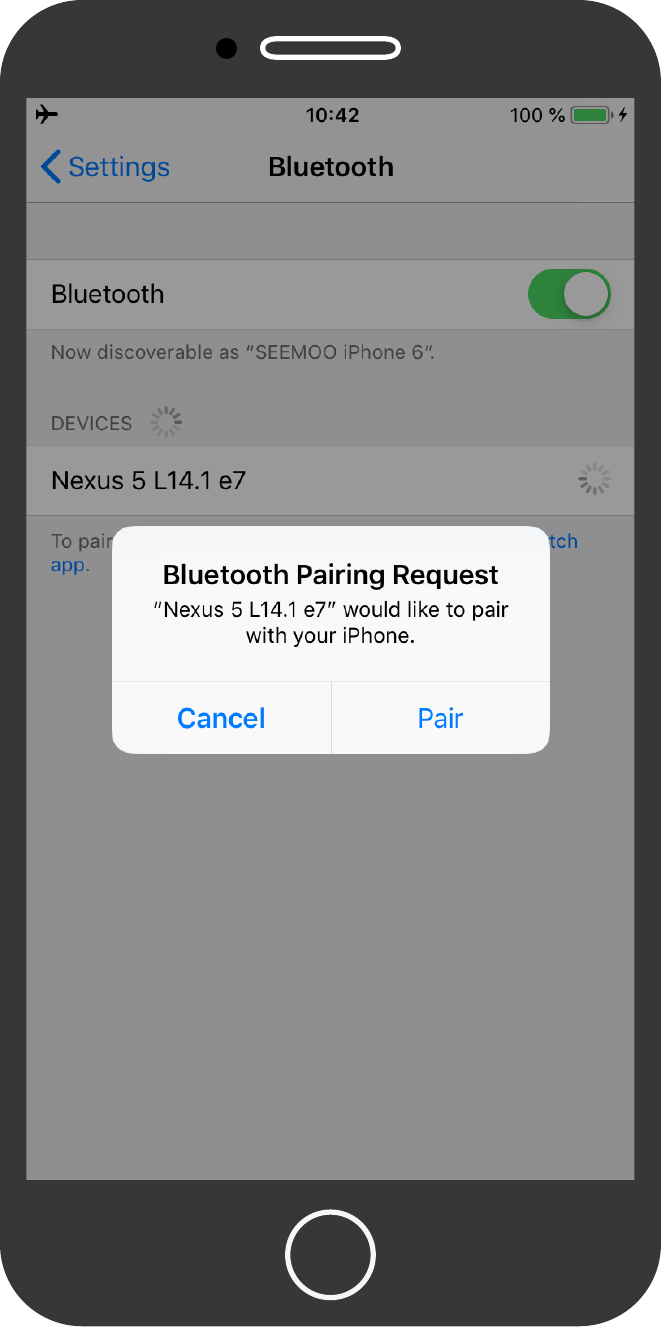}};
    
        \node[inner sep=0pt] (nexus) at (0,0)
    {\includegraphics[height=7.5cm]{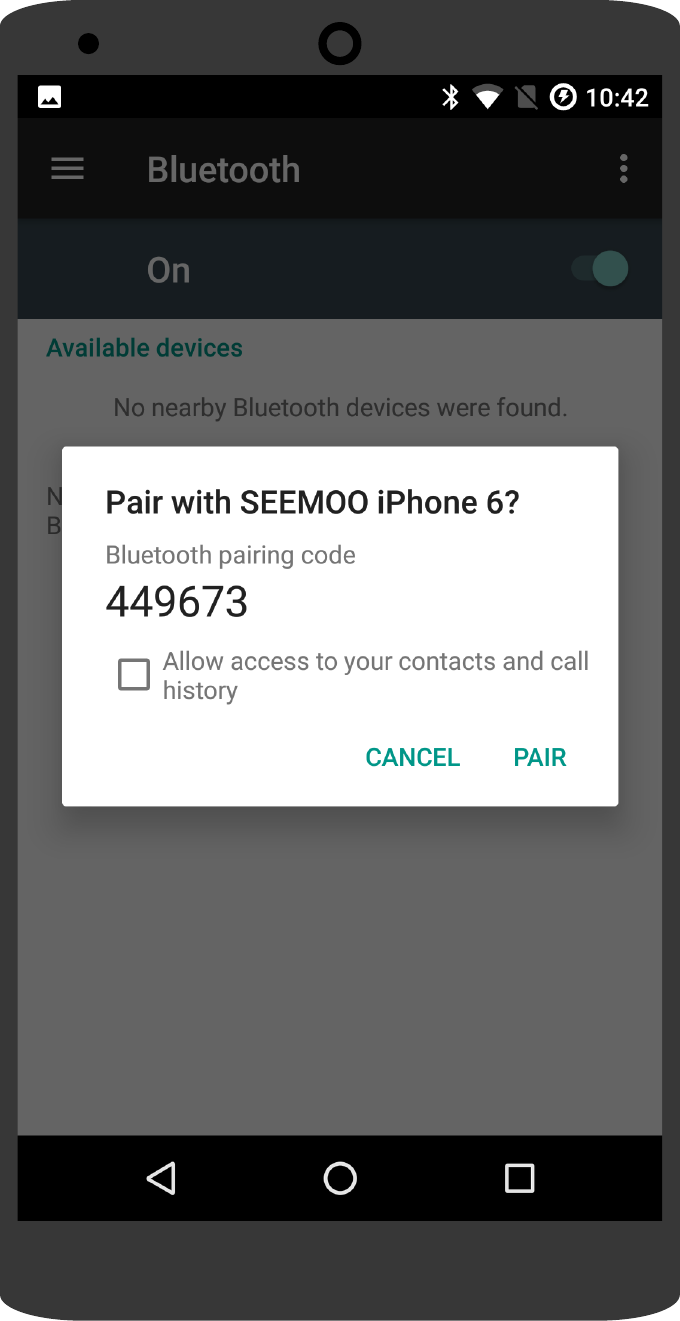}};
    
		\node (n) at (0, -4.1) {\emph{Nexus 5} with IO capabilities};
		\node (i) at (7, -4.1) {\emph{iPhone 6}};
		
		\path[->] (1.92,0) edge node[sloped, anchor=center, above, text width=2.5cm, font=\footnotesize] {Pairing request without IO capabilities} (5.23,0);

	\end{tikzpicture}
	}
	\end{center}
	\caption{``Just Works'' Pairing on \emph{iOS} 12.1.4.}
	\label{fig:nino}
\end{figure}

\autoref{fig:nino} shows a modified \emph{Nexus 5} pairing with an \emph{iPhone 6}. The only modification in the \ac{SSP} state machine function located at \hex{303D4} is to always set the IO capabilities to \hex{03} in variable \hex{20387D}, with \hex{03} resembling no input and no output according to the Bluetooth 5.0 specification~\cite{2016:SIG}. Algorithms within \emph{Nexus 5} are not coded to 
consider the case of no longer having a display with such a change, 
and continue passing a code to \emph{Android}, which actually works because keys are derived similarly either way. \emph{iPhone 6} simply displays a ``yes/no'' message and pairing is achieved. Yet, operating systems seem to cache some of these information locally, making tampering with \ac{IO} capabilities hard if not applied within each pairing. To ensure a \emph{Niño} attack works, an attacker should be present during the first pairing. Users will not be displayed any warnings if a \emph{Niño} attack fails due to detection of inconsistent \ac{IO} capabilities, but pairing will be aborted.



\subsection{ECDH Device Pairing Vulnerability Scan}
\label{ssec:ecdh}

Another application of \mytool is a security test for the fixed coordinate invalid curve attack (CVE-2018-5383)~\cite{2018:biham}.
This attack affects device pairing based on \ac{ECDH}, both \ac{SSP} used in classic Bluetooth and \ac{LE SC} used in \ac{BLE}.
The underlying protocol vulnerability exists if a Bluetooth implementation does not check for invalid curve parameters in the key. Since the Bluetooth standard does not consider countermeasures against this type of attack mandatory, vulnerable implementations are widespread. \todo{Oli: Did the 4.1 standard include this section at all? I honestly only noticed it for the later ones.} Testing for this attack is not straightforward, though, as the authors did not release any tools and the attack requires special hardware.

We implement the semi-passive fixed coordinate invalid curve attack, which zeroes the y-coordinates of both public keys, belonging to the \mytool smartphone and the device under test. If the device under test does not perform an invalid curve parameter check, this semi-passive attack would be successful in 25\% of all device pairing attempts under the original attack's assumptions~\cite{2018:biham}. In contrast, the \mytool success rate is raised to 50\% by setting an even smartphone private key, which is possible since the attack model is not a wireless link under control but the actual smartphone initiating a connection. Although not performing this attack as \ac{MITM}, testing other smartphones is fast and portable with our solution.

\mytool could also help to fix this vulnerability in older \emph{Broadcom} Bluetooth chips. Applying security patches this way is feasible if a vendor is no longer supporting old chips or reacts slowly in publishing updates.



\section{Discovering and Fighting New Vulnerabilities}
\label{sec:new_exploits}

While exploring command handlers we have found a severe vulnerability affecting a variety of devices, as described in \autoref{ssec:exploit}.
 To analyze the vulnerability's impact, we implement runtime testing and emulation solutions, as detailed in \autoref{ssec:exploit_analysis}, which can also be used to detect further bugs or simply to locate and dissect more functions and realize more features. In \autoref{ssec:firewall} we come up with a solution that can help against even more attacks.

\subsection{Remote Code Execution Vulnerability}
\label{ssec:exploit}

%

%
%

In the following we explain how we discovered CVE-2018-19860, a vulnerability allowing to execute a subset of functions within a large set of \emph{Broadcom} firmwares.
The vulnerability only requires the device under attack to have Bluetooth enabled, no previous pairing or visibility to the attacker are required.

\subsubsection{Vulnerability}
While dissecting the \ac{LMP} dispatcher and analyzing its functions, we found vendor specific commands. The Bluetooth standard only defines vendor specific commands on \ac{HCI} layer which can be locally executed on the host as superuser, but \emph{Broadcom} uses LMP opcode \lstinline{0x00} for \ac{BPCS} mapped by \lstinline{lm_BPCS_getLmpInfoType}. Exploring this unusual LMP handler we discovered that it accepts 256 input values in the field following opcode \hex{00}. \emph{Broadcom} only intended to implement \ac{BPCS} commands \hex{0000}--\hex{0005}. Each of these commands is defined in the same handler table syntax as the standard LMP handler described in \autoref{ssec:link_manager}, but because of a missing range check commands \hex{0006}--\hex{00ff} interpret memory located after the intended \ac{BPCS} handler table similarly.

\begin{figure}[b]
	\begin{center}
	\scalebox{0.75}{
	\begin{tikzpicture}[minimum height=0.55cm, node distance=0.9cm]
		\node[text width=4cm] (input) {LMP input: \texttt{00 \textcolor{red!70!black}{95} ...}};
		\draw[-,color=red!70!black] (0.5,-0.2) -- (0.9, -0.2);
		\draw[->,color=red!70!black] (0.7,-0.2) -- (0.7,-0.8) -- (1,-0.8);
		\node[text width=5cm] (bpcs) at (3.6,-0.8) {\textbf{LMP BPCS handler table}};
		\node[text width=7cm] (bpcsf) at (5,-2) { \textcolor{black!20}{\texttt{00}} \texttt{00} Features request\newline  \textcolor{black!20}{\texttt{01}} \texttt{01} Features response \newline \textcolor{white}{\texttt{02 02}} ... \newline  \textcolor{black!20}{\texttt{05}} \texttt{05} BFC accept};
		
		\node[text width=7cm] (other) at (4.6,-3.2) {\textbf{Next (unknown) handler table}};
		\node[text width=5cm] (otherf) at (4.0,-3.9) { \textcolor{red!70!black}{\texttt{06}} \texttt{00} ...\newline \textcolor{white}{\texttt{07 01}} ... };
		
		\node[text width=7cm] (hcif) at (4.6,-5.6) {\textbf{HCI link control handler table \newline HCI link policy handler table \newline HCI host controller handler table \newline HCI info parameter handler table \newline HCI status parameter handler table \newline HCI test handler table}};
		\node[text width=7cm] (hcif) at (5.0,-7.1) { \textcolor{red!70!black}{\texttt{95}} \texttt{03} Enable device under test mode};

		\node[text width=7cm] (hciv) at (4.6,-7.6) {\textbf{HCI vendor specific handler table}};
		\node[text width=7cm] (hcivf) at (5.0,-8.1) { \textcolor{red!70!black}{\texttt{BD}} \texttt{4E} Launch RAM (wrong parameters)};
		
		\draw[->,color=red!70!black] (1.3,-2.62) -- (0.7,-2.62) -- (0.7,-7.1) -- (1.3, -7.1);
		
	\end{tikzpicture}
	}
\end{center}
\caption{Jumping from one handler to another, exemplary handlers picked from \emph{BCM4339} firmware.}
\label{fig:handler_vuln}
\end{figure}
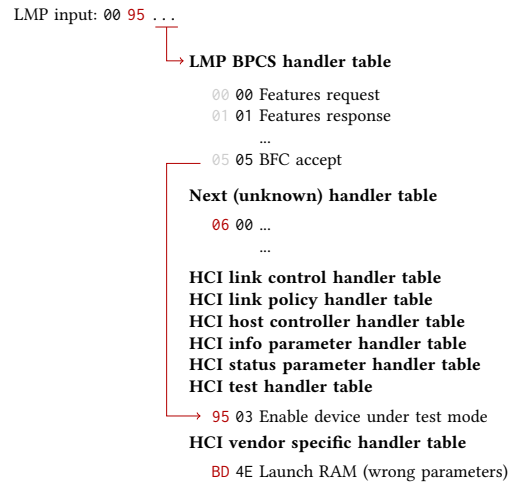

Compilers tend to put similar library functions close to each other. Even though compiling is hard to predict and contents are different for each \emph{Broadcom} chip, it is very likely that handler tables are put after each other. On \emph{Nexus 5} other handler tables are indeed put after the \ac{BPCS} handler table and include an interesting selection of \ac{HCI} commands shown in \autoref{fig:handler_vuln} that never should be launched from a remote host, such as \lstinline{Launch_RAM}. Function arguments are passed in different only partially controllable registers when called over this path, for example, \lstinline{Launch_RAM} is executed but always launches an invalid address. Execution of invalid commands or accessing invalid memory just crashes the Bluetooth firmware and has no further side effects. On some operating systems, the driver restarts Bluetooth automatically within a minute, but even in this case ongoing connections are interrupted. In general, chances to simply crash the \emph{Broadcom} firmware are high.

Interpreting out of bounds commands in \autoref{fig:handler_vuln} also results in extremely long \ac{LMP} packets, i.e. the handler for command \hex{0095} has a length of 219 bytes. When trying to use \mytool's \lstinline{sendlmp} \lstinline{00 95}, the packet seems to be sent according to the monitoring hooks but is never actually sent because the \ac{LMP} packet buffer is limited to 32 bytes. When attacking other smartphones with a \emph{Nexus 5}, the attacker first needs to patch the vulnerable \ac{BPCS} handler locally and can then change length and option fields. A \emph{Nexus 5} under attack will not wait to receive all 219 bytes and just interprets the LMP packet with the wrong length.

\emph{BCM4339} on \emph{Nexus 5} is a combo chip using a shared antenna for Bluetooth and Wi-Fi. Further investigating crashes on \emph{Nexus 5} shows that crashes result in 2--5\,s interrupts on 2.4\,GHz Wi-Fi only. 5\,GHz Wi-Fi is not affected. Similar Wi-Fi interrupts also happen during normal Bluetooth pairing. However, they do not seem to happen when regularly exchanging data. Default configuration of many systems is to restart Bluetooth after a crash, an attacker can reduce Wi-Fi performance multiple times by attacking again when the Bluetooth of the victim reappears. Analyzing possible causes for this behavior is complicated---it could range from incorrect sharing of the antenna up to \emph{Android} specific issues.

\begin{table}[b]
	\caption{Known Vulnerable Devices.}
	\label{tab:devices}
	\small
	\begin{tabular*}{\linewidth}{ @{\extracolsep{\fill}} p{3.7cm} l l l}
	
		Device Name & Chip & Version & SubVersion \\
		\hline		
		MacBook Pro 13'' mid 2012 	& \emph{BCM4331}& 6 & 8859 \\
		iPhone 5 					& \emph{BCM4334} & 7 & 16653 \\
		iPhone 5s					& \emph{BCM4334}& 7 & 8707 \\
		Xperia Z5					& \emph{BCM43xx}	& 7 & 8975 \\
		Nexus 5, Xperia Z3, \newline Samsung Galaxy Note 3						& \emph{BCM4339}	& 7 & 24841 \\
		Raspberry Pi 3				& \emph{BCM43430A1}	& 7 & 8713 \\
		Huawei Honor 8				& \emph{BCM4345} & 8 & 24857 \\
		iPhone 6						& \emph{BCM4345} & 8 & 16649 \\
		MacBook Pro 13'' early 2015 	& 	\emph{BCM2070x}			& 8 & 8609 \\
		MacBook Pro 13'' early 2015 \#2 	& 	\emph{BCM20703A1}			& 8 & 8614 \\
		MacBook Pro 2016	A1707		& 	\emph{BCM20703A2}		& 8 & 8752 \\
		MacBook Pro 2016				& 	\emph{BCM20703A2}			& 8 & 8774 \\

	\end{tabular*}
\end{table}

\subsubsection{Affected Devices}
Vulnerable devices are easy to detect as they do not answer \ac{BPCS} commands with \mbox{\lstinline{LMP_not_accepted}.} Very old \emph{Broadcom} chips do not implement these commands at all and are not vulnerable. 
\emph{Broadcom} did not share a list of vulnerable devices after we reported the issues, but we found an interesting selection listed in \autoref{tab:devices} by testing devices we had access to.
A major version number of 5 or lower seems to be secure against our attack, but corresponds to the outdated Bluetooth 3.0 standard~\cite{linkmgrassignednums}. Only some of the chips with major version 8 were vulnerable, which corresponds to Bluetooth 4.2.

Prior to pairing, devices need to exchange their version number within a \lstinline{LMP_version_req} and \lstinline{LMP_version_resp} to know which version of the Bluetooth standard is supported by each other. This request also includes a subversion that further specifies the firmware version running on the chip. 
This gives the attacker hints which payload could execute meaningful functions.

We extracted firmware from various vulnerable and non-vulnerable devices to further trace down when the function \lstinline{lm_BPCS_getLmpInfoType} was fixed and if handlers located behind are indeed \ac{HCI} or similar interesting functions. Firmware internals of those versions that give most insights about patching are listed below.
\begin{description}
\item[Raspberry Pi 3 / \emph{BCM43430A1}] is vulnerable and its build date is June 2 2014. It has a few callback function tables located in the vulnerable memory region, but most of the memory is filled with encryption constants.
\item[Raspberry Pi 3+ / \emph{BCM4345C0}] is \emph{not} vulnerable and its build date is August 19 2014.
\item[MacBook Pro 2016 / \emph{BCM20703A2}] is vulnerable and its build date is October 22 2015. Various callback and \ac{HCI} function tables are located in the vulnerable memory region.
\end{description}

We assume that Broadcom internally discovered this bug in summer 2014, but was not aware of its criticality at all when adding the missing \ac{BPCS} opcode check. Moreover, a newer build date does not necessarily mean that a firmware has the newest library versions, as the \emph{BCM20703A2} example shows.


As firmware development cycles are quite long---even the evaluation board is shipped with a one year old firmware---devices released in 2016 are still vulnerable.  We estimate around half of the devices with \emph{Broadcom} Bluetooth chips actively used in December 2018 to be vulnerable.

%

\subsection{Firmware Emulation and Fuzzing}
\label{ssec:exploit_analysis}
Exploring abilities gained with the \ac{BPCS} vulnerability is hard. In order to analyze crash causes and impact of non-crashing functions we develop a toolchain that does not only allow hunting exploits but actually exploring flow of information throughout the firmware.

\subsubsection{Tracepoints}
Functions are reached over long call graphs with parameters depending on many dynamic inputs. The \mytool command \lstinline{tp add 0x3f3f4} adds a tracepoint to \lstinline{LMP_dispatcher}. Each tracepoint is realized as \emph{Patchram} hook. Once a tracepoint is called, it is deleted from \emph{Patchram} to prevent the firmware from being stuck in tracepoints while debugging. A tracepoint imitates the behavior of the original \emph{Broadcom} fault handler contained in the firmware and dumps all register contents as well as stack and heap but continues operation afterward. Multiple tracepoints can be observed in a row, i.e. to check behavior of multiple functions called for a given input.

An example output for \lstinline{LMP_dispatcher} is shown in \autoref{lst:tracepoint}. 
The link register points to a function \emph{Broadcom} named \lstinline{lm_handleLmpMsg}, \lstinline{r0} points to \lstinline{lm_curCmd}.

\subsubsection{Emulation with Unicorn and Radare2}
To investigate the LMP handler remote code execution, we use registers from a tracepoint of \lstinline{LMP_dispatcher} but install a second tracepoint in the handler for \ac{BPCS} itself to dump the according stack and heap contents of a \ac{BPCS} payload. With a combination of \emph{Unicorn} and \emph{Radare2} we modify a \emph{Python} script that emulates \lstinline{LMP_dispatcher}~\cite{radare2, unicorn, hugo}. \ac{BPCS} commands are passed by changing the according blocks in the stack and heap memory region. Emulation of an LMP command stops if the end of \lstinline{LMP_dispatcher} is reached or a timeout is exceeded.

\begin{figure}[b]
\begin{lstlisting}[frame=single, basicstyle=\small\ttfamily, caption={Tracepoint in \texttt{LMP\_Dispatcher}.}, label={lst:tracepoint}]
[*] Tracepoint 0x3f3f4 was hit and deactivated:
    pc:  0x0003f3f4	lr:   0x00008c33
    sp:  0x0021734c	cpsr: 0x00000000
    r0:  0x00200478	r1:   0x002179a8
    ...
    r11: 0x00000000	r12:  0x40000000
\end{lstlisting}
\end{figure}

\begin{figure}[b]
\begin{lstlisting}[frame=single, basicstyle=\small\ttfamily, caption={Call Trace Starting at \texttt{LMP\_Dispatcher}.}, label={lst:calltrace}]
0x3f3f4, 0x3f400, 0x42c04, 0x42c0a, 0x3f406,
0x3f40e, 0x3f418, 0x4a868, 0x4a87a, 0x3f41e,
0x3f426, 0x3f44c, 0x3f44e, 0x3f34c, 0xd30d0,
0xd3152, 0xd315e, 0xd3188, 0xd3192, 0xd3130,
0x3f456, 0x3f458, 0x3f464, 0x00000, 
@\textcolor{red!70!black}{invalid memory 0x661e1000}, \textcolor{red!70!black}{invalid memory 0x65f7f3c0}@ 
\end{lstlisting}
\end{figure}

A call trace for \lstinline{LMP_dispatcher} is given in \autoref{lst:calltrace}. In this case, the function is passed the LMP packet \hex{000a}. Following handler tables as in \autoref{fig:handler_vuln}, position \hex{0a} in the misinterpreted \ac{BPCS} handler table simply calls a \lstinline{NULL} pointer that is then interpreted as a function address, and \ac{ROM} contents at \hex{00000} are executed. This trace is still very short because it crashes immediately, however traces can become very complex and hard to analyze.

\subsubsection{Call Trace and Memory Interpretation}
The emulator recognizes branches and extracts addresses when entering a new block after a branch to generate a detailed call trace. Moreover it dumps memory after finishing execution, which can later be compared to the original state.
\ac{BPCS} commands that depend on arguments passed to the handler can be located by emulating the same commands multiple times with different inputs and then searching for traces with different execution paths or by memory differences.

\subsubsection{Crafting Exploits: Turn Remote Devices Into Jammers}
One function standing out in memory analysis changes the content of a very high memory section, which turns out to be responsible for test mode configuration. Further investigation reveals that the LMP payload \lstinline{0x0095} calls the HCI handler function enabling device under test mode. Without being embedded into HCI logic the function gets executed but never appears in \emph{HCI Snoop Log} on the host. Therefore, a host under attack will not be able to filter or observe this event.

\begin{figure}[b]
\begin{lstlisting}[frame=single, basicstyle=\small\ttfamily, caption={Remote Jammer Attack.}, label={lst:lmp_jammer}]
@\textcolor{gray}{\# LMP\_set\_AFH to disable hopping}@
sendlmp 60 0000000000ffffffffffffffff0000
@\textcolor{gray}{\# Enable test mode via exploit}@
sendlmp 00 95     
@\textcolor{gray}{\# LMP\_test\_activate}@
sendlmp 56 00     
@\textcolor{gray}{\# LMP\_test\_control, TX frequency 2433 MHz}@
sendlmp 57 545575755555555255 
\end{lstlisting}
\end{figure}

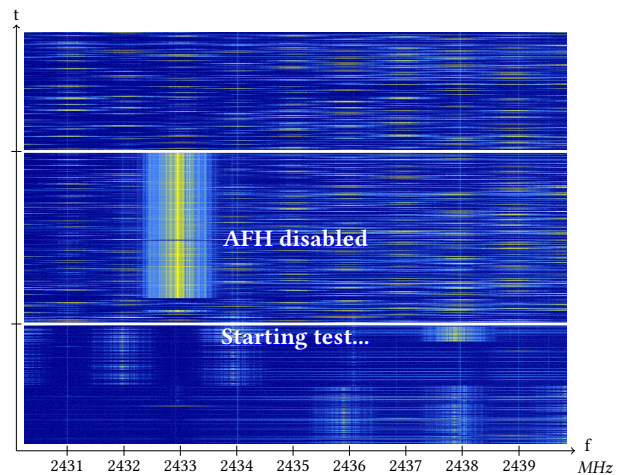
\begin{figure}
	\begin{center}
	\scalebox{0.75}{
    \begin{tikzpicture}[minimum height=0.55cm, node distance=0.7cm, scale=0.9]

        \node[inner sep=0pt] (testmode) at (0,0)
    {\includegraphics[width=0.54\textwidth]{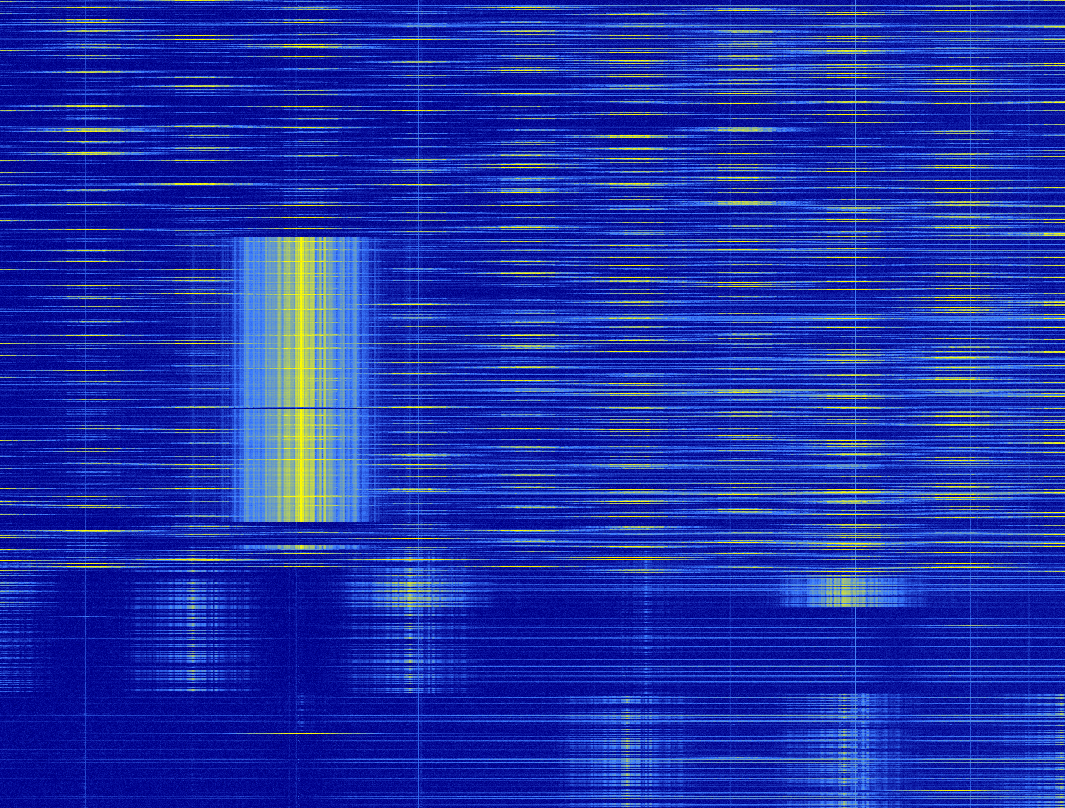}};

		\draw[->] (-5.5,-4.2) -- (5.5,-4.2);
		\node[text width=1cm, align=center] (f) at (5.9, -4.3) {\ \ \ f \newline\emph{MHz}};
		\draw[->] (-5.5,-4.2) -- (-5.5,4.2);
		\node (t) at (-5.5, 4.4) {t};

		\draw[-,line width=0.5mm,color=white] (-5.4,-1.7) -- (5.5,-1.7);
		\draw[-,line width=0.5mm,color=white] (-5.4,1.7) -- (5.5,1.7);
		\draw[-] (-5.4,-1.7) -- (-5.6,-1.7);
		\draw[-] (-5.4,1.7) -- (-5.6,1.7);
		
		\draw[-] (-4.5,-4.1) -- (-4.5,-4.3);
		\draw[-] (-3.38,-4.1) -- (-3.38,-4.3);
		\draw[-] (-2.27,-4.1) -- (-2.27,-4.3);
		\draw[-] (-1.16,-4.1) -- (-1.16,-4.3);
		\draw[-] (-0.05,-4.1) -- (-0.05,-4.3);
		\draw[-] (1.06,-4.1) -- (1.06,-4.3);
		\draw[-] (2.17,-4.1) -- (2.17,-4.3);
		\draw[-] (3.28,-4.1) -- (3.28,-4.3);
		\draw[-] (4.4,-4.1) -- (4.4,-4.3);
		
		\node (f1) at (-4.5,-4.5) {2431};
		\node (f2) at (-3.38,-4.5) {2432};
		\node (f3) at (-2.27,-4.5) {2433};
		\node (f4) at (-1.16,-4.5) {2434};
		\node (f5) at (-0.05,-4.5) {2435};
		\node (f6) at (1.06,-4.5) {2436};
		\node (f7) at (2.17,-4.5) {2437};
		\node (f8) at (3.28,-4.5) {2438};
		\node (f9) at (4.4,-4.5) {2439};

		\node[color=blue, font=\fontsize{12}{0}\selectfont] (startx) at (0.015,-2.015) {\textbf{Starting test...}};
		\node[color=white, font=\fontsize{12}{0}\selectfont] (start) at (0,-2) {\textbf{Starting test...}};
		\node[color=blue, font=\fontsize{12}{0}\selectfont] (afhx) at (0.015,-0.015) {\textbf{AFH disabled}};
		\node[color=white, font=\fontsize{12}{0}\selectfont] (afh) at (0,0) {\textbf{AFH disabled}};

	\end{tikzpicture}
	}
	\end{center}
	\caption{Device Under Test Mode Exploit.}
	\label{fig:lmp_jammer_afh}
	\vspace{-2em} 
\end{figure}

After enabling test mode with the malicious payload, \lstinline{LMP_test_activate} and \lstinline{LMP_test_control} can be used to run tests. The master controls test mode and the slave is the device under test. Each test is performed for multiple seconds. During a test, master and slave usually hop over all channels defined in their \ac{AFH} configuration.
 In conjunction with \lstinline{LMP_set_AFH}, \ac{AFH} can be disabled on the attacked device; thus, causing it to jam a selected frequency, which could be pilots of a Wi-Fi signal the attacker wants to block. To enable jamming of a single frequency,
the \lstinline{LMP_set_AFH} payload must contain a valid Bluetooth clock, which requires writing an assembly patch. A sequence of frames successfully turning a \emph{Nexus 5} and \emph{Xperia Z3} into a remote jammer is given in \autoref{lst:lmp_jammer}. Note that \lstinline{LMP_set_AFH} is not under perfect control of the attacker without passing a proper Bluetooth clock. \ac{AFH} is not stopped on the attacking device but the device under attack nevertheless temporarily accepts the \ac{AFH} configuration. With the presented payloads, the device under attack keeps hopping on all Bluetooth channels most of the time as shown in  \autoref{fig:lmp_jammer_afh}.

\subsection{MAC Address Filter}
\label{ssec:firewall}
As a basic defense against devices injecting malicious frames, a \ac{MAC} address filter can be used.
Typically, users only use a few devices they trust such as a headset.
We implement a whitelist inside \mbox{\lstinline{LMP_dispatcher}.}
Untrusted devices are rejected within \ac{LMP}---an attacker not imitating a valid \ac{MAC} address 
cannot establish connections and not even tamper with any of the \ac{LMP} handlers or protocols above.
To successfully imitate a \ac{MAC} address, attackers need to analyze traffic with a \ac{SDR} for a while to calculate the target address from demodulated signal chunks~\cite{2009:ossmann}. In contrast to Wi-Fi, Bluetooth chunks do not contain full \ac{MAC} addresses, therefore guessing these is harder than in Wi-Fi.

Currently, the \ac{MAC} address filter is implemented as a simple \emph{Python} proof of concept executed at runtime by using \mytool as a library. It can be integrated into more user friendly solutions, such as an \emph{Android} app that automatically generates a permanent HCD file allowing only connections from devices that were successfully paired at the time of its generation.

A \ac{MAC} address filter is not required to defend against CVE-2018-19860 by now, as most vendors rolled out patches. \emph{Broadcom} did not include us in the feedback loop after informing vendors in December 2018, but we tested previously affected devices and can confirm that our vulnerability was fixed in \emph{iOS} 12.1.3.
The \emph{macOS} update fixing the vulnerability increases the \ac{LMP} subversion by one.

%
%

\section{Related Work}
\label{sec:related}
%
%
%
%

To the best of our knowledge, \mytool is the only framework that allows Bluetooth lower layer functionality to be patched on off-the-shelf devices. \mytool was inspired by \emph{Nexmon}, a similar project from our group, which realizes modifications on the Wi-Fi part of \emph{BCM4339}~\cite{2018:schulz}. Despite residing in the same chip, Wi-Fi patching is realized very differently. 

External tools for Bluetooth analysis always use an \ac{MITM} setup. No changes are required on the devices to be sniffed. However, the \ac{MITM} must successfully follow frequency hopping and overhear initial pairing to break encryption.
Older variants of classic Bluetooth and \ac{BLE} have a weak encryption and passive \ac{MITM} is sufficient. If appropriate encryption is used, the \ac{MITM} must actively inject traffic during key exchange. Depending on the physical setup and tool implementation, breaking pairing might take multiple tries---in either case the setup is rather artificial and users will be most likely aware of sniffing. \mytool does not have any of these issues in a sniffing setup and supports classic Bluetooth as well as \ac{BLE}, since it is part of the communication, but due to serving a similar purpose we briefly introduce other Bluetooth sniffing tools in the following.

The most popular open-source solution \emph{Ubertooth}~\cite{ubertooth:project} is an \ac{SDR} extended with Bluetooth capabilities, therefore providing \ac{PHY} access if frequency hopping succeeds. \ac{BLE} sniffing support is quite good as its \ac{PHY} is rather simple, but classic Bluetooth sniffing is very limited and not even attacks on bad encryption techniques are implemented~\cite{spill2007bluesniff}. At a cost of 120\$ it is comparably cheap~\cite{ubertooth_attify}. It might be possible that \mytool can also modify \ac{PHY} behavior. The \emph{Nexmon} project altered the Wi-Fi part of \emph{BCM4339} and actually found a re-programmable real-time \emph{D11} core~\cite{2018:schulz}.

Due to the simplicity of \ac{BLE}, there exist a lot of low-price sniffers, such as the \ac{BLE} 4.0 \emph{Bluefruit LE Sniffer} for \$25~\cite{bluefruit}.

The \ac{BLE} 4.2 \emph{CY5677} platform~\cite{cy5677}, which was used to implement the \ac{ECDH} device pairing attack~\cite{2018:biham}, only costs \$20 and comes with CySmart to re-program the USB dongle's behavior~\cite{cysmart}. For lower layers, it only allows to set some variables, which includes scan, connection, and security parameters. Advanced protocol support is only provided for layers above \ac{HCI}, such as \ac{GATT} and \ac{L2CAP}.

\emph{Ellisys} offers the best available Bluetooth tools in their Tracker, Explorer and Vanguard series~\cite{ellisys-bv1}, but at the highest price (around \$10k-\$20k). According to the Vanguard documentation, live views and logs of all layers are displayed, and injections on \ac{HCI} layer are possible. Though, there is no interface to implement custom lower-layer functions.


\section{Discussion}
\label{sec:discussion}

After describing the implementation \mytool and demonstrating that it is possible to develop
capable tools with the framework, we discuss its strengths,  limitations, and impact.


\paragraph{Portability to Other Platforms}
An ongoing task is porting \mytool to other members of the \emph{Broadcom} and \emph{Cypress} Bluetooth-Wi-Fi chip series.
We already ported some of the features to the \emph{BCM4358} chip, which is part of the \emph{Nexus 6P}, \emph{Samsung Galaxy S6} and \emph{Samsung Galaxy S6 edge}.
We verified that reading and writing \ac{RAM} works on very old chips, such as the \emph{BCM2070} with a firmware from 2008.
Moreover, instead of requiring a rooted \emph{Android} device with a recompiled kernel module, we developed more a user-friendly solution based on \emph{BlueZ} sockets on \emph{Linux} to support the cheaper  \emph{Raspberry Pi 3/3+} (\emph{BCM4343/BCM4345}).
Another work in progress is firmware analysis of the Bluetooth 5.0~\cite{2016:SIG} compatible chips such as \emph{CYW20735}, which still provide \emph{Broadcom} commands for firmware modification.
Since newer Bluetooth standards are backward compatible, recent \emph{Broadcom} firmware still contains many functions from older versions. Nevertheless, changes in the underlying operating system and data structures exist; hence, porting still requires reverse engineering efforts.

To port \mytool to non-\emph{Broadcom} platforms, their firmware update mechanisms have to be reverse engineered. Significant functions and data structures have to be located similar to our description in \autoref{sec:reversing}. Doing so is very time consuming and \emph{Broadcom} Bluetooth chips have a market share in the range of 20\%--30\%. Thus, we did not make any efforts to support other vendors so far.

\paragraph{Early Bluetooth 5.1 Adaptation}
\mytool can also help to implement features from the new standard before they are available on the market. The newest Bluetooth 5.1~\cite{2019:SIG} was just released and as of now no hardware is available.
Bluetooth 5.1 adds \ac{AoA} and \ac{AoD} localization. As these require \ac{PHY} adjustments, successful reverse engineering down to a level where \emph{Broadcom} chips can be used as \ac{SDR} is required.
However, Bluetooth 5.1 also introduces new advertisement and power saving mechanisms for the link layer, which are probably easy to implement as binary patches.

%

\paragraph{Stability}
When experimenting with \emph{Ubertooth}, getting a setup that actually sniffs a complete connection is challenging. This problem class does not apply to \mytool---one of the bigger advantages of being able to modify a stable and well-maintained firmware while it is part of a fully
functional \emph{Android} Bluetooth stack.
Yet, any crash due to
firmware modifications will cause the \emph{Android} system to reset the chip; therefore it is recommended to carefully test all \mytool patches to keep the same level of stability as in the original firmware.
Stability depends on finding a memory region that is not used by other threads, which is very challenging on the \emph{Nexus 5} but easy on the Bluetooth 5.0 evaluation kit. We are currently improving stability by reverse engineering how memory management and message queues work.


\paragraph{Wi-Fi and Bluetooth Coexistence}
Most \emph{Broadcom} chips support Bluetooth and Wi-Fi. They run on different ARM cores but share some components such as the 2.4\,GHz antenna. We found that \emph{Broadcom} supports interesting features such as a \ac{RSSI} sweep over all Bluetooth channels on the local interface to figure out how they interfere with Wi-Fi and select channels properly. Artifacts in the evaluation kit firmware even indicate that LTE coexistence management exists. How well-separated Bluetooth, Wi-Fi and LTE components are in terms of performance and security remains an open research topic.

\paragraph{Wireless Firmware Security}
Patching the vulnerable \ac{LMP} handler requires only 14 Bytes of machine code on \emph{Nexus 5}. This appears to be a simple fix, but installing it to all vulnerable devices is not. Smartphones in a non-rooted standard configuration cannot use \mytool and need to be provided with operating system updates that include new HCD files. Older vulnerable devices might not be updated---\emph{Broadcom} confirmed us in December 2018 that they will provide patches, but they did not mention which devices will be fixed specifically.

Publishing patches as HCD files is generally risky, as they are neither encrypted nor signed by \emph{Broadcom}. Therefore, patches are easy to reverse engineer and releasing a patch will disclose the vulnerability. Moreover, information leaks such as a full list of function and object names in \lstinline{patch.elf} of \emph{WICED Studio} are dangerous for \emph{Broadcom}'s security by obscurity approach.

After first reverse engineering results of \emph{Broadcom} Wi-Fi chips were published~\cite{nexmon:project} it did not take long until security researchers found severe security bugs in its implementation~\cite{2017:googleprojectzero}. They further analyzed these bugs to ultimately obtain privilege escalation in the operating system with the Bluetooth driver's superuser rights on \emph{Apple} devices.
Even proprietary protocol extensions such as \ac{AWDL} based on Wi-Fi and Bluetooth had major security issues in the past~\cite{milan}.
Reverse engineering shows that security in today's wireless chips depends a lot on obscurity, which is not a sufficient measure.

In general, security-concerned users should turn off their device's Wi-Fi and Bluetooth whenever they are not needed. Modern smartphones sometimes require to disable Wi-Fi and Bluetooth in the system settings to not just have them in a standby mode.
On the one hand, this also increases battery lifetime and avoids tracking of users. On the other hand, a modern smartphone without wireless services is severely impaired in its abilities.


\paragraph{Bluetooth Security Evaluation Toolkit}
\mytool and its applications are filling the gap in the inventory of security-related Bluetooth tools.
Not only important and novel features such as \ac{LMP} monitoring
and injection are added, but also the \ac{ECDH} device pairing vulnerability scan demonstrates its usage as testing toolkit for advanced attacks. \mytool is limited to run on \emph{Broadcom} and \emph{Cypress} chips due to its code modification mechanisms, but devices under test can be of any kind.


Indeed, \mytool is powerful for \ac{LMP} analysis. 
By leveraging \ac{LMP} injection
capabilities it is possible to audit other Bluetooth devices by enumerating proposed pairing and encryption
capabilities.

Analyzing the extracted firmware enables emulation for fuzzing as well as static code analysis. This paper only analyzed \ac{LMP} but there are more Bluetooth subprotocols that can contain bugs or are of interest for developing and testing optimizations.


\section{Conclusion}
\label{sec:conclusion}

%
%
%
%
%

In the motivation we pointed out the shortage of openly available research tools that allow for efficient experimenting on Bluetooth protocol layers beneath the \ac{HCI}.
This work documents the development of \mytool, a novel analysis
and research platform for lower Bluetooth protocol layers. The framework is
based on multiple \emph{Broadcom} Bluetooth chipsets, whose firmware has been reverse engineered and documented in the process.
Undocumented patching mechanisms to even change ROM contents could be dissected from the firmware update
procedure and therefore \mytool benefits from the ability to reuse and modify the
existing Bluetooth stack implemented in the controller.
These patching mechanisms exist at least throughout firmware versions from 2008 to 2018. We found that newly released off-the-shelf devices typically are shipped with two year old wireless chips, so even if \emph{Broadcom} decided to secure their patching mechanisms today, \mytool will continue to work on hardware released over the next few years.

Eventually, the framework has proven itself as a powerful tool for \ac{LMP} monitoring
and injection, as well as a security testing toolkit. 
With this paper, we lay the groundwork for the community to further contribute to advanced Bluetooth research. Even though \mytool is based on \emph{Broadcom} chips, it can be used to probe other, non-\emph{Broadcom} Bluetooth devices.


Overall, the addition of a new tool to the inventory of  researchers
advances the state of Bluetooth security and facilitates integration of new protocol features.
Furthermore, reverse engineering the firmware of a commercial Bluetooth
controller paves the way for open reviews by the security community and may
reveal vulnerabilities that have been hidden up to this point.


\section*{Acknowledgments}
\ifblinded
Acknowledgments blinded for review.
\else
We thank \emph{Broadcom} for providing CVE-2018-19860 patches to vendors. We also thank Oliver Pöllny for proofreading and feedback.

This work has been funded by
the \textsc{dfg}
as part of project S1
within \textsc{sfb 1119 crossing}, as part of project C.1 within the \textsc{rtg 2050} ``Privacy and Trust for Mobile Users'', \textsc{sfb 1053 maki}; and
the \textsc{bmbf} and the State of Hesse within \textsc{crisp-da}.

\fi

\balance

\bibliographystyle{abbrv} 
\bibliography{bibliographies}


\end{document}